\documentclass[12pt]{article} 
\usepackage[top=20mm,bottom=30mm, left=20mm, right=20mm]{geometry}
\usepackage{graphicx,amsmath,amssymb,url,enumerate,mathrsfs,epsfig,color,cite}
\numberwithin{equation}{section}

\newtheorem{prop}{Proposition}[section]
\newtheorem{thm}{Theorem}[section]

\newtheorem{lem}{Lemma}[section]
\newtheorem{rem}{\it Remark}[section]
\title{Material homogeneity and strain compatibility in thin elastic shells}
\author{Ayan Roychowdhury and Anurag Gupta}
\date{}
\begin{document}
\maketitle

\begin{abstract}
We discuss several issues regarding material homogeneity and strain compatibility for materially uniform thin elastic shells from the viewpoint of a 3-dimensional theory, with small thickness, as well as a 2-dimensional Cosserat theory. A relationship between inhomogeneity and incompatibility measures under the two descriptions is developed. More specifically, we obtain explicit forms of intrinsic dislocation density tensors characterising inhomogeneity of a dislocated Cosserat shell. We also formulate a system of governing equations for the residual stress field emerging out of strain incompatibilities which in turn are related to inhomogeneities. The equations are simplified for several cases under the Kirchhoff-Love assumption.
\end{abstract}



\section{Introduction}
In this paper we explore the notions of material homogeneity and strain compatibility in materially uniform thin elastic shells. In general terms, a materially uniform body is said to be homogeneous if there exists a globally differentiable map from any configuration of the body to its undistorted state \cite{noll, epsbook}; this usually amounts to the body being free of topological defects. If a body is materially homogeneous then the associated strain field is compatible; the converse, however, is not true. Whenever material inhomogeneity leads to incompatibility, it subsequently becomes a source for internal stresses in the body \cite{kroner81}. A fundamental problem in micromechanics is, for a given distribution of defects in a solid, to determine the resulting state of deformation and stress field. The motivation for the present work is to develop a framework where this problem can be addressed for a broad class of thin structures.

Homogeneity in materially uniform bodies has been explored both in 3-dimensional (3-d) solids \cite{kondo2, noll, bilby, kroner81, epsbook} and 2-dimensional (2-d) structured solids like shells \cite{ericksen, wangshell, edeleon, eps1, zubov1, zubov2, steigmann14} but, unlike the former, appropriate inhomogeneity measures (or defect densities) and their relation with strain incompatibility has not been sufficiently developed for the latter. On the other hand, although strain compatibility relations for non-linear shells have appeared in the literature for over five decades \cite{koiter, reissner, malcolm, eps1, ciar1}, inclusion of incompatibilities has been attempted only recently \cite{efrati, derezin}. Toward these ends one can model an inhomogeneous shell either as a 3-d body with small thickness, thereby using the available infrastructure of 3-d inhomogeneity theories, or alternatively as a 2-d Cosserat surface. It is well established that the 2-d approach is amenable to analytical and numerical computations as well as for physical applications. It will therefore be useful to formulate inhomogeneity measures and incompatibility relations in terms of intrinsic surface quantities. Our methodology is to use previously established results from the 3-d theory to derive appropriate relations both in terms of 3-d as well as 2-d variables. In doing so, we need to define a 3-d metric field in terms of intrinsic 2-d strain measures; this definition essentially captures the geometry of the shell theory that we wish to work with, see Remark \ref{metricstrain} for further details. A related notion of embedded homogeneity in thin structures, keeping in mind this dual characterization, has been recently discussed in the context of beams \cite{epsarc1,epsarc2}. Here, for a given a 1-dimensional materially inhomogeneous beam, the nature of material homogeneity of the 2-d solid, that it is actually made up of, is investigated.

In Section 2, we revisit the classical fundamental theorem of Riemannian geometry in a new light, such that it is applicable to the  3-d continuum theory of topological defects. Several remarks are provided to clarify the usefulness of the theorem for material homogeneity, strain compatibility, and evaluation of residual stresses. Most of these results are well known both in differential geometry and in continuum theory of defects \cite{ciar1, noll, kroner81}. In Section 3, we apply the results from Section 2 to explore the issue of homogeneity and compatibility in thin elastic shells. In order to use the 3-d results, we construct a 3-d metric field using intrinsic strain measures of a Cosserat shell. More specifically, we derive explicit relationships between 3-d continuous dislocation distribution and its 2-d analogue on a Cosserat surface as well as representations of the latter in terms of Cosserat kinematical variables. We therefore obtain a complete characterization of the dislocation density distribution on a Cosserat shell. We discuss strain compatibility for shells, again within both 3-d and 2-d frameworks, and formulate the complete set of incompatibility fields while emphasizing their role in determining the residual stress field in the shell. We also derive the relationship of the incompatibility fields with intrinsic dislocation densities associated with the shell. In Section 4, we restrict our attention to Kirchhoff-Love shells and derive governing equations for residual stress determination under further simplifications. In particular, we show that under Kirchhoff-Love constraint a dislocated shell can support only in-surface dislocations.

\section{Homogeneity and compatibility in a 3-d elastic solid}
Let $\mathcal{B}$ be a simply-connected open set in $\mathbb{R}^3$ whose closure $\bar{\mathcal{B}}$ has a piecewise smooth boundary; moreover, let $\mathcal{B}$ be such that it can be covered with a single chart. Hence, $\mathcal{B}$ admits a global parametrization $\boldsymbol{X}:(\theta^1,\theta^2,\theta^3=:\zeta)\in\mathbb{R}^3\to\mathcal{B}$. The notation $(\cdot)_{,i}$ is a shorthand for the partial derivative $\frac{\partial(\cdot)}{\partial \theta^i}$. Although one can put various geometric structures (e.g., a connection, a metric) on $\mathcal{B}$, it naturally inherits the Euclidean structure of $\mathbb{R}^3$ including the Euclidean inner product (denoted by $\cdot$). Let $\boldsymbol{G}_i:=\boldsymbol{X}_{,i}(\theta^\alpha,\zeta)$, $G_{ij}:=\boldsymbol{G}_i\cdot\boldsymbol{G}_j$, $[G^{ij}]:=[G_{ij}]^{-1}$, and $\boldsymbol{G}^i:= G^{ij}\boldsymbol{G}_j$. The Roman indices vary between $1$ and $3$.

We have the following
\begin{thm}\label{thm:1}
Let $L^p_{ij}$ be sufficiently smooth real functions on $\mathcal{B}$, satisfying
\begin{equation}
R^i_{jkl}:= L^i_{jl,k}- L^i_{jk,l}+L^h_{jl} L^i_{hk}-L^h_{jk} L^i_{hl}=0\,\,\,\textrm{on}\,\,\mathcal{B}.
\label{hyp:1}
\end{equation}

(i) Then, there exists a sufficiently smooth invertible matrix field (denoted by $[\hat H_{ij}(\theta^\alpha,\zeta)]$) on $\mathcal{B}$ such that 
\begin{equation}
L^q_{ij}=(\hat H^{-1})^{ql}\,\hat H_{li,j}.
\label{form:1}
\end{equation}

(ii) Moreover, let $[g_{ij}]$ be a positive definite symmetric matrix field on $\mathcal{B}$ satisfying
\begin{equation}
 g_{ij;k}:=g_{ij,k}-L^p_{ik} g_{pj}-L^p_{jk} g_{pi}=0\,\,\,\textrm{on}\,\,\mathcal{B}.
 \label{hyp:2}
\end{equation}
Then, \begin{equation}
       g_{ij}=\boldsymbol{g}_i\cdot\boldsymbol{g}_j,
       \label{form:2}
      \end{equation}
 where $\boldsymbol{g}_j:=\hat H_{ij}\boldsymbol{G}^i$.
 
(iii) Furthermore, if $L^p_{ij}=L^p_{ji}$, then there exists a sufficiently smooth diffeomorphism $\boldsymbol{\chi}:=\chi_i\boldsymbol{G}^i$ on $\mathcal{B}$ such that 
\begin{equation}
g_{ij}=\chi_{p|i}\chi_{q|j}G^{pq},
\label{form:3}
\end{equation}
where $\chi_{p|i}:=\chi_{p,i}-\Gamma^j_{pi}\chi_j$, with $\Gamma^j_{pi}:=\frac{1}{2}g^{jn}(g_{np,i}+g_{ni,p}-g_{pi,n})$. Here, $[g^{ij}]$ is the inverse of $[g_{ij}]$.
\end{thm}

This well-known result is a variant of the classical fundamental theorem of Riemannian geometry (Theorem 1.6-1 in \cite{ciar1}) where otherwise the symmetry  of $L^p_{ij}$ is assumed \textit{a priori}. The importance of asymmetric $L^p_{ij}$ in describing continuous distribution of defects (or material inhomogeneities) in a 3-d continuous solid body is discussed below. Several additional remarks are made before proceeding to the proof, so as to elaborate the utility of this theorem in describing the geometry of defects. In the next section, we will use this theorem to discuss analogous issues in a 2-d structured solid.

\begin{rem}
{\rm (Materially uniform elastic solid) 
The trivial manifold $\mathcal{B}$ is our prototype for the theory of a continuous material body. The elements in $\mathcal{B}$ are called material points. The material structure of the body is modelled through a constitutive response function which can be used to understand the geometric nature of defects in the body. In the present article, we will assume the body to be simple hyper-elastic solid, for which the constitutive response function is given by a positive definite mapping $\hat W: Sym^+\times\mathcal{B}\to\mathbb{R}$. Here, $Sym^+$ denotes the set of real, symmetric, positive definite matrices. In addition, the body is assumed to be materially uniform in the sense that there exist another positive definite mapping $W: Sym^+\to\mathbb{R}$ and a matrix field $[\hat H_{ij}]$ over $\mathcal{B}$ with $det[\hat H_{ij}]>0$, such that 
\begin{equation}
\hat W([A_{ij}],\boldsymbol{X})=W([\hat H_{pi}(\boldsymbol{X})][A_{pq}][\hat H_{qj}(\boldsymbol{X})])
\label{uniformity}
\end{equation}
is satisfied for all $[A_{ij}]\in Sym^+$ and all $\boldsymbol{X}\in\mathcal{B}$ \cite{noll}. The field $[\hat H_{ij}]$ is known as the material uniformity field and is, in general, not unique.}
\end{rem}

\begin{rem}
{\rm (The material space) Let the manifold $\mathcal{B}$ be equipped with an affine connection with coefficients $L^p_{ij}$ and a metric field with components $g_{ij}$. The nature of the connection $L^p_{ij}$, known as the \textit{material connection}, and the metric $g_{ij}$, the \textit{material metric}, is informed by the underlying material structure of the body in the following way. The geometry of the \textit{material space}, defined as the triple $(\mathcal{B};\,L^p_{ij},g_{ij})$, brings out the defective nature of the material body. The Riemann-Christoffel curvature of the material connection, as defined by \eqref{hyp:1}$_1$, is a measure of the disclination content of the body. A zero disclination density (equation \eqref{hyp:1}) is tantamount to the existence of a distant parallelism in the material space. This translates into the existence of well-defined vector fields $\boldsymbol{g}_j:=\hat H_{ij}\boldsymbol{G}^i$ ($[H_{ij}]$ is the material uniformity field) which are covariantly constant with respect to the material structure (see \eqref{ap:3}), such that the material connection is necessarily given by \eqref{form:1}; $\boldsymbol{g}_i$ are known as the \textit{material uniformity} or \textit{crystallographic bases}. On the other hand, the material metric $g_{ij}$ is derived from the usual Euclidean metric of the embedding space $\mathbb{R}^3$ as a pull-back of the material uniformity field; it is always well-defined for a solid body \cite{noll}. The non-metricity associated with the material space, defined by the covariant derivative of the material metric with respect to the material connection (see \eqref{hyp:2}$_1$), represents the presence of metric anomalies, such as point defects or thermal strains, in the material body. Under zero non-metricity (equation \eqref{hyp:2}), the metric is necessarily related to the crystallographic bases as in \eqref{form:2}. The material space is, in general, non-Riemannian because the torsion tensor of the material connection, which has the components
\begin{equation}
T^p_{ij}:=\frac{1}{2}(L^p_{ij}-L^p_{ji}),
\end{equation}
is not necessarily zero. The third order skew tensor $T^p_{ij}$ (or, equivalently, its second order axial tensor, with components $\alpha^{kp}:=\frac{1}{2}\varepsilon^{ijk}T^p_{ij}$) provides a measure for the density of dislocation-like anomalies within the material body. The body is called \textit{materially homogeneous} if and only if the Riemann-Christoffel curvature tensor, the non-metricity tensor, and the torsion tensor associated with the material space vanish identically at all points.}
\end{rem}

\begin{rem}
{\rm (The Riemannian space)
The metric structure itself gives rise to another affine connection on $\mathcal{B}$, i.e. the Levi-Civita connection with coefficients
\begin{equation}
\Gamma^p_{ij}:=\frac{1}{2}g^{pq}(g_{qj,i}+ g_{qi,j}-g_{ij,q}),
\label{ap:levivicita}
\end{equation}
which is, by definition, torsion free. It has however a non-zero Riemann-Christoffel curvature
\begin{equation}
 K^i_{jkl}:=\Gamma^i_{jl,k}- \Gamma^i_{jk,l}+\Gamma^h_{jl} \Gamma^i_{hk}-\Gamma^h_{jk} \Gamma^i_{hl},
\end{equation}
which provides a measure for the \textit{incompatibility} in the elastic Lagrangian strain field $E_{ij}:=\frac{1}{2}(g_{ij}-G_{ij})$. Thus, there exists a \textit{Riemannian space} $(\mathcal{B};\,g_{ij})$\footnote{In a Riemannian space, metric $g_{ij}$ determines all the geometric structures.} associated with the material body.}
\end{rem}

\begin{rem}
{\rm (Strain compatibility) Assume $R^i_{jkl} = 0$ and $ g_{ij;k}=0$; hence the body is possibly dislocated. 
The material connection can then be shown to be related to the Levi-Civita connection as \cite{noll}
\begin{equation}
L^p_{ij}=\Gamma^p_{ij}+C^p_{ij},
\end{equation}
where
\begin{equation}
C^p_{ij}:= T^p_{ij}-g^{mp}(T^q_{mj}\,g_{iq}+T^q_{mi}\,g_{jq})
\end{equation}
are the components of the contorsion tensor of the material connection $L^p_{ij}$. Moreover, the respective curvatures are related to each other:
\begin{equation}
R^i_{jkl}=K^i_{jkl}+C^i_{jl|k}-C^i_{jk|l}+C^h_{jl} C^i_{hk}-C^h_{jk} C^i_{hl},
\label{stress}
\end{equation}
where the subscript $_|$ denotes covariant derivative with respect to the connection $\Gamma^p_{ij}$. With $R^i_{jkl}=0$, the last relation provides a non-linear PDE for the metric for a given dislocation density field. For a vanishing dislocation density, $K^i_{jkl} = 0$ which is the compatibility relation for 3-d nonlinear elasticity. The inverse is however not true.  A solution of the PDE \eqref{stress} with $R^i_{jkl}=0$ and $K^i_{jkl}=0$ is $C^p_{ij}=(Q)^{lp}Q_{lj,i}$ where $[Q_{ij}]$ is a sufficiently smooth orthogonal matrix field over $\mathcal{B}$ (several other non-trivial solutions are given in \cite{arash}). When this is the case, the elastic strain is compatible even when the material is no longer homogeneous, in the sense that $T^p_{ij}\ne 0$. Such a state of the material is known as \textit{contorted aeolotropy} \cite{noll}. For an isotropic elastic body in a state of contorted aeolotropy, however, vanishing of the curvature $K^i_{jkl}$ of its Riemannian space implies material homogeneity; this is due to the non-uniqueness in the torsion tensor for bodies with continuous symmetry groups \cite{noll}. Consequently, $K^i_{jkl}$ should be taken as a genuine measure of material inhomogeneity for an isotropic solid.}
\end{rem}

\begin{rem}
{\rm (The residual stress field) The material space $(\mathcal{B};\,L^p_{ij},g_{ij})$ is the \textit{relaxed} or stress-free state of the material body. Due to its general non-Riemannian nature, originating from the presence of material defects, it is often not physically realizable as a connected set in $\mathbb{R}^3$. This in general leads to an incompatibility elastic strain field. For an elastic solid, in  the absence of external forces and displacement boundary conditions, incompatibility of strain is the only source for non-trivial stress field \cite{kroner81}.  
 Equation \eqref{stress}, with $R^i_{jkl}=0$, describes how dislocations as a source of material inhomogeneity yield strain incompatibility inside the material body and, together with the equation of motion and boundary conditions, form the governing equations for determining the stress field.}
\end{rem}


We need the following lemma to prove Theorem \ref{thm:1} (for a proof see Proposition 11.36 in \cite{lee}).
\begin{lem}\label{lem:1}
 Let $V$ be a simply connected open set in $\mathbb{R}^n$ and let $[A_{ij}(\theta^k,z^k)]$ be a  sufficiently smooth $(n\times n)$ matrix field on $V\times\mathbb{R}^n$, for $(\theta^k,z^k)\in V\times \mathbb{R}^n$. If
 \begin{equation}
  \frac{\partial A_{ij}}{\partial \theta^k} +A_{pk} \frac{\partial A_{ij}}{\partial z^p}=\frac{\partial A_{ik}}{\partial \theta^j} +A_{pj} \frac{\partial A_{ik}}{\partial z^p}
 \end{equation}
on $V\times \mathbb{R}^n$, then given any $(\theta^1_0,\ldots,\theta^n_0)\in V$ and $(z^1_0,\ldots,z^n_0)\in \mathbb{R}^n$, there exist unique smooth maps $f_i:V\to \mathbb{R}$, for $i=1,\ldots,n$, such that 
\begin{subequations}
\begin{align}
& \frac{\partial f_j}{\partial \theta^i} = A_{ji}(\theta^l,f_k)\,\,\,\textrm{on}\,\,V,\\
 &f_k(\theta^1_0,\ldots,\theta^n_0)=z^k_0.
 \end{align}
\end{subequations}
\end{lem}


\noindent\textbf{Proof of Theorem 2.1:} $(i)$ Let $[\hat H^0_{lj}]$ be an invertible matrix and let us consider the following partial differential equation
\begin{subequations}
\begin{align}
 \hat H_{lj,i}(\boldsymbol{X})&=L^p_{ji}(\boldsymbol{X}) \hat H_{lp}(\boldsymbol{X}),\,\,\forall\boldsymbol{X}\in\mathcal{B},\\
 \hat H_{lj}(\boldsymbol{X}_0)&=\hat H^0_{lj},
\end{align}
\label{ap:1}%
\end{subequations}
for some generic $\boldsymbol{X}_0\in\mathcal{B}$. For each integer $l=1,2,3$, let $f_j:=\hat H_{lj}$ and $f^0_j:=\hat H^0_{lj}$. Then, the above system of PDEs yield
\begin{subequations}
\begin{align}
  f_{j,i}(\boldsymbol{X})&=L^p_{ji}(\boldsymbol{X}) f_{p}(\boldsymbol{X}),\,\,\forall\boldsymbol{X}\in\mathcal{B},\\
  f_{j}(\boldsymbol{X}_0)&= f^0_{j}.
\end{align}
\label{ap:1'}%
\end{subequations}
The integrability condition of the above system is given by \eqref{hyp:1}, which follows from Lemma \ref{lem:1}, where $A_{ij}(\theta^k,f_m):=L^m_{ij}(\theta^k)f_m$. Equation \eqref{ap:1} can also be written in the form \eqref{form:1}.

$(ii)$ Clearly, the vector fields $\boldsymbol{g}_j:=\hat H_{lj}\boldsymbol{G}^l$ satisfy the following problem by definition:
\begin{subequations}
\begin{align}
 &\boldsymbol{g}_{j;i}(\boldsymbol{X}):=\boldsymbol{g}_{j,i}(\boldsymbol{X})-L^p_{ji}(\boldsymbol{X})\boldsymbol{g}_p(\boldsymbol{X})=\boldsymbol{0},\,\,\boldsymbol{X}\in\mathcal{B},\\
&\boldsymbol{g}_j(\boldsymbol{X}_0)=\boldsymbol{g}^0_j,
\end{align}
\label{ap:3}%
\end{subequations}
 where $\boldsymbol{g}^0_j:= H^0_{ij}\boldsymbol{G}^i(\boldsymbol{X}_0)$. 
Hence, the matrix field $[\boldsymbol{g}_i \cdot \boldsymbol{g}_j]$ satisfies
\begin{subequations}
\begin{align}
&(\boldsymbol{g}_i \cdot \boldsymbol{g}_j)_{,k}=L^p_{ik}\, (\boldsymbol{g}_p \cdot \boldsymbol{g}_j) + L^p_{jk}\, (\boldsymbol{g}_p \cdot \boldsymbol{g}_i)\,\,\,\textrm{in}\,\,\mathcal{B} ,\\
&(\boldsymbol{g}_i \cdot \boldsymbol{g}_j)(\boldsymbol{X}_0)=g^0_{ij},
\end{align}
\label{ap:2}%
\end{subequations}
where $g^0_{lj}:=\hat H^0_{pl}\hat H^0_{qj}G^{pq}(\boldsymbol{X}_0)$. The partial differential equation for $g_{ij}$ given in the hypothesis~\eqref{hyp:2}, along with the condition $g_{ij}(\boldsymbol{X}_0)=g^0_{ij}$, is identical to the problem~\eqref{ap:2}. The solution to the problem is however unique, see e.g. Theorem 1.6-1 in \cite{ciar1}. Hence, $g_{ij}=\boldsymbol{g}_i \cdot \boldsymbol{g}_j\,\,\,\textrm{in}\,\,\mathcal{B}$.

$(iii)$ When $L^p_{ij}=L^p_{ji}$, the solution $\hat H_{ij}$ to Equations \eqref{ap:1} satisfies
\[\hat H_{li,j}-\hat H_{lj,i}=\hat H_{li|j}-\hat H_{lj|i}=0.\]
For a simply connected domain $\mathcal{B}$, such that the curvature associated with $\Gamma^p_{ij}$ is zero, Poincar\'e's lemma implies that, for each integer $l$ (with values $1,2,$ and $3$), there exists a sufficiently smooth $\chi_l: \mathcal{B}\to\mathbb{R}$ such that
\[\hat H_{li}=\chi_{l|i}\,\,\,\textrm{in}\,\,\mathcal{B},\]
i.e., we have the existence of a smooth enough diffeomorphism $\boldsymbol{\chi}$ that satisfies \eqref{form:3}.\hfill$\square$


\section{Homogeneity and compatibility of thin elastic shells}
A shell is a 3-d solid body whose one of the dimensions is much smaller than the other two. Alternatively a shell can be described as a 2-d structured (Cosserat) solid. Our aim is to revisit Theorem \ref{thm:1}, and the associated remarks, for a shell like body under certain assumptions on the nature of the deformation and the mechanical behaviour. We expect the resulting insights to be of value in studying distribution of defects, and the resulting stress field, in thin structures. Let us adapt the embedded curvilinear coordinates $(\theta^1,\theta^2,\zeta)$ in $\mathcal{B}$ such that the coordinates $(\theta^1,\theta^2)$ lie along the orientable mid-surface $\omega$ and $\zeta$ along the normal direction to $\omega$. Then, $\mathcal{B}$ can be parametrized as
\begin{equation}
 \boldsymbol{X}(\theta^\alpha,\zeta)=\boldsymbol{R}(\theta^\alpha)+\zeta \boldsymbol{N}(\theta^\alpha),
 \label{current}
\end{equation}
where $\boldsymbol{R}(\theta^\alpha)$ is the parametrization of $\omega$, $\boldsymbol{N}(\theta^\alpha)$ is the unit normal field on $\omega$, and $\zeta\in[-h,h]$, where $2h$ is the thickness of the shell assumed to be constant. The Greek indices take a value of either $1$ or $2$. Let $\boldsymbol{A}_\alpha:= \boldsymbol{R}_{,\alpha}$. The first and second fundamental form of $\omega$ are $A_{\alpha\beta}:=\boldsymbol{A}_\alpha\cdot\boldsymbol{A}_\beta$ and $B_{\alpha\beta}:=-\boldsymbol{N}_{,\beta}\cdot\boldsymbol{A}_\alpha$, respectively. Let $[A^{\alpha\beta}]:=[A_{\alpha\beta}]^{-1}$ and $\boldsymbol{A}^\alpha:= A^{\alpha\beta}\boldsymbol{A}_\beta$. With respect to the notation in the previous section, we have
\begin{enumerate}[(i)]
\item $\boldsymbol{G}_\alpha(\theta^\alpha,\zeta)=(\delta^\beta_{\alpha}-\zeta B^{\beta}_{\alpha})\boldsymbol{A}_{\beta}(\theta^\alpha)$ and  $\boldsymbol{G}_3(\theta^\alpha,\zeta)=\boldsymbol{N}(\theta^\alpha)$, where $B^\beta_\alpha:=A^{\beta\gamma}B_{\alpha\gamma}$,
\item $G_{\alpha\beta}=A_{\alpha\beta}-2\zeta B_{\alpha\beta}+\zeta^2 C_{\alpha\beta}$ and $G_{i 3}=\delta_{i3}$,  where $C_{\alpha\beta}:= B_\alpha^\tau B_{\tau\beta}$,
\item $G^{\alpha\beta}=A^{\alpha\beta}+2\zeta A^{\alpha\mu} A^{\beta\nu} B_{\mu\nu}+3\zeta^2 A^{\alpha\mu} A^{\beta\nu} C_{\mu\nu} + O(\zeta^3)$ and $G^{i3}=\delta_{i3}$, and
\item $\boldsymbol{G}^\alpha=\boldsymbol{A}^\alpha+\zeta\,B^\alpha_\beta\,\boldsymbol{A}^\beta + \zeta^2\,C^\alpha_\beta\,\boldsymbol{A}^\beta+ O(\zeta^3)$ and $\boldsymbol{G}^3=\boldsymbol{N}$, where $C^\alpha_\beta:= A^{\alpha\tau} C_{\tau\beta}$.
 \end{enumerate}
 Here, following Landau's notation, for $\boldsymbol{f}:\mathbb{R}\to\mathbb{R}^k$, we write $\boldsymbol{f}(s)=O(s)$ if and only if there exist positive constants $M$ and $\delta$ such that $||\boldsymbol{f}(s)||_{\mathbb{R}^k}\le M|s|$ for all $|s|<\delta$, and $\boldsymbol{f}(s)=o(s)$ if and only if $\lim_{s\to 0}\frac{||\boldsymbol{f}||_{\mathbb{R}^k}(s)}{s}=0$.

\subsection{Material homogeneity}
A motivation for our study is to derive intrinsic defect density (or inhomogeneity) measures for a 2-d structured body and obtain the required PDEs to solve for stresses and deformation of the inhomogeneous shell. Towards this end we start by considering the following set of strain-like smooth fields on $\omega$: $E_{\alpha\beta}=E_{\beta\alpha}$ such that $[A_{\alpha\beta}]+2[E_{\alpha\beta}]$ is positive definite, $\Delta_\alpha$, $\Delta(\ne -1)$, $\Lambda_{\alpha\beta}$, and $\Lambda_\alpha$. The choice of these strain-measures, whose physical nature is remarked below, dictates the kinematical nature of the thin structure. For instance, the straining of a membrane can be described in terms of $E_{\alpha\beta}$ only. 

\begin{thm}\label{thm:2}
(i) Let $L^p_{ij}$ be sufficiently smooth real functions defined on $\mathcal{B}=\{\boldsymbol{X}(\theta^\alpha,\zeta)=\boldsymbol{R}(\theta^\alpha)+\zeta \boldsymbol{N}(\theta^\alpha)\subset\mathbb{R}^3\}$, $\zeta\in[-h,h]$, satisfying
\begin{equation}
R^i_{jkl}:= L^i_{jl,k}- L^i_{jk,l}+L^h_{jl} L^i_{hk}-L^h_{jk} L^i_{hl}=0\,\,\,\textrm{on}\,\,\mathcal{B}.
\label{hyp:3}
\end{equation}
  Let $[g_{ij}]$ be a symmetric matrix field on $\mathcal{B}$, defined by
\begin{equation}
g_{\alpha\beta}:= a_{\alpha\beta}+\zeta\,P_{\alpha\beta} +\zeta^2 \,Q_{\alpha\beta},\,\,\,g_{\alpha 3}:= \Delta_\alpha+\zeta\,U_\alpha,\,\,\,g_{33}:= V,
\label{metric:defn}
\end{equation}
where
\begin{subequations}
\begin{align}
a_{\alpha\beta} &:= A_{\alpha\beta}+2 E_{\alpha\beta},\\
 P_{\alpha\beta} &:= 2(\Lambda_{(\alpha\beta)}-B_{\alpha\beta}),\\
Q_{\alpha\beta} &:= a^{\sigma\gamma}\,(\Lambda_{\sigma\alpha}-B_{\sigma\alpha})(\Lambda_{\gamma\beta}-B_{\gamma\beta})+\Lambda_\alpha\,\Lambda_\beta,\\
 U_\alpha &:= a^{\sigma\gamma}\,\Delta_\sigma (\Lambda_{\gamma\alpha}-B_{\gamma\alpha})+\Lambda_\alpha (\Delta+1),~\text{and}\\
 V &:= a^{\alpha\beta}\,\Delta_\alpha \Delta_\beta+(\Delta+1)^2,
 \end{align}
\end{subequations}
with $[a^{\alpha\beta}]:=[a_{\alpha\beta}]^{-1}$, such that\footnote{Note that, since $\omega$ is bounded and $[g_{ij}]$ is continuous in $\theta^\alpha$ and $\zeta$, $[g_{ij}]$ will be positive definite on $\mathcal{B}:=\omega\times(-h,h)$ for sufficiently small $h$. Our result is valid for this sufficiently small $h$ and we start with an $\mathcal{B}$ such that $h$ conforms with this small value. For a technical discussion on the issue of smallness of $h$ and positive definiteness of $[g_{ij}]$, please refer to the proof of Theorem 2.8-1 in \cite{ciar1}.}
\begin{equation}
g_{ij;k}:=g_{ij,k}-L^p_{ik} g_{pj}-L^p_{jk} g_{pi}=0\,\,\,\textrm{on}\,\,\mathcal{B}.
 \label{hyp:4}
\end{equation}
Then there exist sufficiently smooth vector fields $\boldsymbol{a}_{\alpha}(\theta^\alpha)$ with $\boldsymbol{a}_1(\theta^\alpha)\times\boldsymbol{a}_2(\theta^\alpha)\ne \boldsymbol{0}$, $\boldsymbol{D}_\alpha(\theta^\alpha)$, and $\boldsymbol{d}(\theta^\alpha)$ with $\boldsymbol{d}(\theta^\alpha)\cdot\boldsymbol{a}_1(\theta^\alpha)\times\boldsymbol{a}_2(\theta^\alpha)\ne 0$, such that (here $\boldsymbol{g}_i$ should be interpreted as introduced in the previous section)
\begin{subequations}
\begin{align}
\boldsymbol{g}_\alpha(\theta^\alpha,\zeta)&=\boldsymbol{a}_\alpha(\theta^\alpha)+\zeta\boldsymbol{D}_\alpha ~\text{and}\\
\boldsymbol{g}_3(\theta^\alpha,\zeta)&=\boldsymbol{d}(\theta^\alpha).
\end{align}
\label{field:link}%
\end{subequations}
In particular,
\begin{subequations}
\begin{align}
\boldsymbol{a}_\alpha(\theta^\alpha)\cdot\boldsymbol{a}_\beta(\theta^\alpha)&=a_{\alpha\beta}(\theta^\alpha),\\
\boldsymbol{D}_\alpha(\theta^\alpha)&=(\Lambda_{\sigma\alpha}-B_{\sigma\alpha})\boldsymbol{a}^\sigma+\Lambda_\alpha\boldsymbol{n},~\text{and}\\
\boldsymbol{d}(\theta^\alpha)&=\Delta_\sigma\boldsymbol{a}^\sigma+(\Delta+1)\boldsymbol{n},
\end{align}
\label{fields}%
\end{subequations}
where $\boldsymbol{a}^\sigma(\theta^\alpha):=a^{\sigma\beta}\boldsymbol{a}_\beta$ and $\boldsymbol{n}(\theta^\alpha):=\displaystyle\frac{\boldsymbol{a}_1(\theta^\alpha)\times\boldsymbol{a}_2(\theta^\alpha)}{|\boldsymbol{a}_1(\theta^\alpha)\times\boldsymbol{a}_2(\theta^\alpha)|}$.

(ii) Furthermore, if the torsion tensor evaluated at the mid-surface vanishes, i.e. $T^p_{ij}(\theta^\alpha,0)=0$, then there exists a sufficiently smooth diffeomorphic image $\hat\omega$ of $\omega$, parametrized by $\boldsymbol{r}(\theta^\alpha)$, such that $\boldsymbol{a}_\alpha(\theta^\alpha)=\boldsymbol{r}_{,\alpha}(\theta^\alpha)$, and $\boldsymbol{D}_\alpha(\theta^\alpha)=\boldsymbol{d}_{,\alpha}(\theta^\alpha)$. In particular, there exists an open  sufficiently smooth diffeomorphic image $\hat{\mathcal{B}}$ of $\mathcal{B}$, parametrized by $\boldsymbol{\chi}(\theta^\alpha,\zeta)=\boldsymbol{r}(\theta^\alpha)+\zeta \boldsymbol{d}(\theta^\alpha)$, such that $\boldsymbol{g}_i(\theta^\alpha,\zeta)=\boldsymbol{\chi}_{,i}(\theta^\alpha,\zeta)$.
\end{thm}
This result is central to our understanding of the defective nature of the shell. We would like to emphasize that the hypothesis \eqref{metric:defn} for the metric structure on the 3-d body manifold $\mathcal{B}$ of the shell is crucial to our proof as well as for interpretation of the results. Relation \eqref{metric:defn} is a generalization of a similar 3-d metric used in the proof of the fundamental theorem (Theorem 2.8-1 in \cite{ciar1}) of embedded surfaces in $\mathbb{R}^3$. Before proving the theorem, we provide several remarks to bring out the importance of this result from the view point of defect mechanics.

\begin{rem}
{\rm (Shell kinematics) Consider a sufficiently thin shell (i.e. ${h}/{R}<<1$, with $R=$ minimum principal radius of curvature of the shell mid-surface for a given deformation) made up of an hyper-elastic solid whose material structure is characterized by a material connection $L^p_{ij}$ and a metric $g_{ij}$, such that the curvature associated with $L^p_{ij}$ is zero (see \eqref{hyp:3}) and the metric is covariantly constant with respect to the material connection (see \eqref{hyp:4}). Therefore the shell material has no intrinsic disclination and metric anomalies. Furthermore if this shell is free of dislocations, i.e. if $T^p_{ij}(\theta^\alpha,\zeta)=0$, it will have a coherent relaxed configuration characterized by a diffeomorphism $\boldsymbol{\chi}$ on $\mathcal{B}$. The notion of sufficient thinness of the shell is then manifested in the particular action of $\boldsymbol{\chi}$ which
maps
\begin{equation}
\mathcal{B}=\{\boldsymbol{X}(\theta^\alpha,\zeta)=\boldsymbol{R}(\theta^\alpha)+\zeta \boldsymbol{N}(\theta^\alpha)\}
\label{shell:current}
\end{equation}
onto the relaxed state
\begin{equation}
 \hat{\mathcal{B}}=\boldsymbol{\chi}(\mathcal{B})=\{\boldsymbol{x}(\theta^\alpha,\zeta)=\boldsymbol{r}(\theta^\alpha)+\zeta \boldsymbol{d}(\theta^\alpha)\}.
 \label{shell:relaxed}
\end{equation}
Here, the stress relaxation process (i.e. the elastic deformation) respects the classical Green-Naghdi approximation for sufficiently thin shells \cite{greennaghdi}. The elastic deformation $\boldsymbol{\chi}:\mathcal{B}\to\hat{\mathcal{B}}$ of a materially homogeneous sufficiently thin shell preserves the `fibrous' structure of $\mathcal{B}$, in the sense that straight transverse sections remain straight and transverse throughout the deformation. Clearly, there are two separate modes of deformation at play: the mid-surface deformation 
\begin{equation}
\boldsymbol{R}(\theta^\alpha)\mapsto\boldsymbol{r}(\theta^\alpha)=\boldsymbol{R}(\theta^\alpha)+\boldsymbol{u}(\theta^\alpha),\label{deform1}
\end{equation}
where $\boldsymbol{u}(\theta^\alpha)$ is a well-defined displacement field on $\omega$, and the director deformation 
\begin{equation}
\boldsymbol{N}(\theta^\alpha)\mapsto\boldsymbol{d}(\theta^\alpha)=\boldsymbol{Q}(\theta^\alpha)\boldsymbol{N}(\theta^\alpha),
\label{deform2}
 \end{equation}
 where $\boldsymbol{Q}(\theta^\alpha):=\boldsymbol{d}(\theta^\alpha)\otimes\boldsymbol{N}(\theta^\alpha)$ is a second-order tensor field defined on $\omega$.
 
Thus, while as a full-fledged 3-d material body the elastic deformation of a materially homogeneous shell is characterized by the diffeomorphism $\boldsymbol{X}(\theta^\alpha,\zeta)\mapsto\boldsymbol{x}(\theta^\alpha,\zeta)$, due to its `fibrous' nature---a mathematical artifact brought forth by the representations \eqref{shell:current} and \eqref{shell:relaxed}---the same elastic deformation is equivalently characterized by the set consisting of a diffeomorphism $\boldsymbol{R}(\theta^\alpha)\mapsto\boldsymbol{r}(\theta^\alpha)$ of the base manifold $\omega$ and a linear isomorphism $\boldsymbol{N}(\theta^\alpha)\mapsto\boldsymbol{d}(\theta^\alpha)$ of the transverse fibres (the director fields) attached to the base manifold. The latter characterization is a manifestation of the Cosserat structure of the shell described by a vector bundle consisting of a 2-d base manifold $\omega$ and an isomorphic copy of a frame of $\mathbb{R}^3$ attached to each point of $\boldsymbol{R}\in\omega$, such that two vectors of the frame at $\boldsymbol{R}$ span $T_{\boldsymbol{R}}\omega$ and the third vector is transverse to it. In the Cosserat picture, the elastic deformation of a materially homogeneous shell is a principal bundle isomorphism of this vector bundle which subsumes a diffeomorphism of the base manifold and a linear isomorphism of the frame field, as just described.\footnote{A point worthwhile  overemphasizing is that unlike the inherently Cosserat-type materials (also known as polar media or materials with microstructures) e.g. liquid crystals, magnetic materials etc., a structural shell (as is presently the case) is not inherently a Cosserat material but merely a mathematical artifact originated from the representations \eqref{shell:current} and \eqref{shell:relaxed}.}
}
\end{rem}
 
\begin{rem}
{\rm (Strain measures)
 The dual characterization of shell kinematics, as outlined above, motivates the particular form \eqref{metric:defn} of the metric structure on the 3-d shell manifold $\mathcal{B}$, relating the two equivalent descriptions of the shell material space. The definition \eqref{metric:defn} is central to all the subsequent claims made in Theorem \ref{thm:2}. We are considering a single director geometrically nonlinear shell theory, taking into account the transverse shear and normal deformation. The fields $a_{\alpha\beta}$, $\Lambda_{\alpha\beta}$, $\Lambda_\alpha$ $\Delta_\alpha$ and $\Delta$ constitute the 2-d metric structure on the shell's \textit{Cosserat material space}. `Sufficient thinness' of the shell is encoded in the definition \eqref{metric:defn} which describes how the 2-d strain fields can be used to construct a 3-d metric on $\mathcal{B}$. The resulting 3-d metric is second-order in the transverse coordinate $\zeta$; it is the unique generalization of the metric 
 \begin{equation}
  g_{ij}(\theta^\alpha,\zeta) = a_{\alpha\beta} - 2\zeta b_{\alpha\beta}+ \zeta^2 a^{\mu\nu}b_{\mu\alpha}b_{\nu\beta}
 \end{equation}
defined in the proof of Theorem 2.8-1 in \cite{ciar1} that considers a Kirchhoff-Love shell model (i.e., $\Lambda_\alpha=\Delta_\alpha=\Delta=0$), where $b_{\alpha\beta}:=-\Lambda_{\alpha\beta}+B_{\alpha\beta}$.
Due to the specific form \eqref{metric:defn} of $g_{ij}$, the material uniformity bases $\boldsymbol{g}_i$ on the 3-d material space $(\mathcal{B};\,L^p_{ij},g_{ij})$ decompose into the form \eqref{field:link}, guaranteeing the existence of the vector fields $\boldsymbol{a}_\alpha$, $\boldsymbol{d}$ and $\boldsymbol{D}_\alpha$ that constitute the \textit{material uniformity bases} on the \textit{Cosserat material space} of the shell. The relations \eqref{fields}, which are analogous to Theorem \ref{thm:1}$(ii)$, relate the metric structure $\{a_{\alpha\beta}, \Lambda_{\alpha\beta}, \Lambda_\alpha, \Delta_\alpha, \Delta\}$ to the material structure $\{\boldsymbol{a}_\alpha,\boldsymbol{D}_\alpha,\boldsymbol{d}\}$ in the Cosserat material space of the shell. The last two relations in \eqref{fields} can be more explicitly written as
  \begin{subequations}
 \begin{align}
 \Delta_{\alpha}&=\boldsymbol{d}\cdot\boldsymbol{a}_\alpha-\boldsymbol{N}\cdot\boldsymbol{A}_\alpha,\\
\Delta&=\boldsymbol{d}\cdot\boldsymbol{n}-\boldsymbol{N}\cdot\boldsymbol{N},\\
\Lambda_{\alpha\beta}&=\boldsymbol{D}_{\beta}\cdot\boldsymbol{a}_\alpha-\boldsymbol{N}_{,\beta}\cdot\boldsymbol{A}_\alpha,~\text{and}\\
\Lambda_{\alpha}&= \boldsymbol{D}_{\alpha}\cdot\boldsymbol{n}-\boldsymbol{N}_{,\alpha}\cdot\boldsymbol{N}.
\end{align}
\end{subequations}
These expressions clearly bring out the strain-like nature of the fields appearing on the left hand side.}
\label{metricstrain}
\end{rem}

\begin{rem}
{\rm (Material inhomogeneity measures) Let us consider the 3-d shell to be free of disclinations and metric defects. According to Theorem \ref{thm:2}$(ii)$, for the kind of shell presently considered (characterized by a metric of the form \eqref{metric:defn}), the material inhomogeneity of the shell is completely determined by the restriction of the 3-d torsion tensor to the mid-surface, i.e. $T^p_{ij}(\theta^\alpha,0)$. This tensor field can be alternatively expressed in terms of the 2-d inhomogeneity measures of the shell:
 \begin{subequations}
\begin{align}
T^\mu_{\alpha\beta}(\theta^\alpha,0)&=(\hat H^{-1})^{\mu\sigma}(\theta^\alpha,0)\mathbb{T}_{\mu\alpha\beta}(\theta^\alpha)+(\hat H^{-1})^{\mu 3}(\theta^\alpha,0)\mathbb{T}_{3\alpha\beta}(\theta^\alpha),\\
T^3_{\alpha\beta}(\theta^\alpha,0)&=(\hat H^{-1})^{3\sigma}(\theta^\alpha,0)\mathbb{T}_{\mu\alpha\beta}(\theta^\alpha)+(\hat H^{-1})^{33}(\theta^\alpha,0)\mathbb{T}_{3\alpha\beta}(\theta^\alpha),\\
T^\mu_{\alpha 3}(\theta^\alpha,0)&=(\hat H^{-1})^{\mu\sigma}(\theta^\alpha,0)\mathbb{T}_{\mu\alpha3}(\theta^\alpha)+(\hat H^{-1})^{\mu 3}(\theta^\alpha,0)\mathbb{T}_{3\alpha3}(\theta^\alpha),~\text{and}\\
T^3_{\alpha 3}(\theta^\alpha,0)&=(\hat H^{-1})^{3\sigma}(\theta^\alpha,0)\mathbb{T}_{\mu\alpha3}(\theta^\alpha)+(\hat H^{-1})^{33}(\theta^\alpha,0)\mathbb{T}_{3\alpha3}(\theta^\alpha),~\text{where}
\end{align}
\end{subequations}
\begin{subequations}
 \begin{align}
 \mathbb{T}_{\mu\alpha\beta}(\theta^\alpha):=\hat{H}_{\mu[\alpha,\beta]}(\theta^\alpha,0) &= H_{\mu[\alpha,\beta]},\\
 \mathbb{T}_{3\alpha\beta}(\theta^\alpha):=\hat{H}_{3[\alpha,\beta]}(\theta^\alpha,0) &= H_{3[\alpha,\beta]},\\
 \mathbb{T}_{\mu \alpha3}(\theta^\alpha):=\hat{H}_{\mu[\alpha,3]}(\theta^\alpha,0) &= F_{\mu3,\alpha}-F_{\mu\alpha}+B^\nu_\mu H_{\nu\alpha},~\text{and}\\
 \mathbb{T}_{3\alpha3}(\theta^\alpha):=\hat{H}_{3[\alpha,3]}(\theta^\alpha,0) &= F_{33,\alpha}-F_{3\alpha},
 \end{align}
 \end{subequations}
are the 2-d Cosserat inhomogeneity measures, with $H_{\alpha\beta}(\theta^\alpha):=\boldsymbol{A}_\alpha \cdot \boldsymbol{a}_\beta$, $H_{3\alpha}(\theta^\alpha):=\boldsymbol{N}\cdot \boldsymbol{a}_\alpha$, $F_{\alpha\beta}:=\boldsymbol{A}_\alpha\cdot\boldsymbol{D}_\beta$, $F_{3\beta}:=\boldsymbol{N}\cdot\boldsymbol{D}_\beta$, $F_{\alpha3}:=\boldsymbol{A}_\alpha\cdot\boldsymbol{d}$ and $F_{33}:=\boldsymbol{N}\cdot\boldsymbol{d}$. In the above expressions a square bracket in the subscript indicates the skew part of the field with respect to the enclosed indices (on the other hand, round brackets are used to indicate the symmetric part). The vanishing of 2-d Cosserat inhomogeneity measures is equivalent to  $T^p_{ij}(\theta^\alpha,0)=0$. We note the following interpretations:
 \begin{enumerate}
  \item The component $\mathbb{T}_{\mu\alpha\beta}(\theta^\alpha)$ measures in-surface dislocation density. Its  vanishing implies the existence of the in-surface components $r_{\alpha}(\theta^\alpha):=\boldsymbol{r}\cdot\boldsymbol{A}_{\alpha}$ of the surface diffeomorphism $\boldsymbol{\chi}(\theta^\alpha,0)$.
  \item The component $\mathbb{T}_{3\alpha\beta}(\theta^\alpha)$ measures the out-of-surface dislocation density. Its vanishing implies the existence of the out-of-surface component $r_{3}(\theta^\alpha):=\boldsymbol{r}\cdot\boldsymbol{N}$ of the surface diffeomorphism $\boldsymbol{\chi}(\theta^\alpha,0)$.
  \item The component $\mathbb{T}_{\mu\alpha 3}(\theta^\alpha)$ measures the in-surface integrability of the director field $\boldsymbol{d}$. Its vanishing implies $(\boldsymbol{D}_{\alpha}-\boldsymbol{d}_{,\alpha})\cdot\boldsymbol{a}_\mu=0$.
  \item The component $\mathbb{T}_{3\alpha 3}(\theta^\alpha)$ measures the out-of-surface integrability of the director field $\boldsymbol{d}$. Its vanishing  implies $(\boldsymbol{D}_{\alpha}-\boldsymbol{d}_{,\alpha})\cdot\boldsymbol{n}=0$.
 \end{enumerate}
 
In the Cosserat picture the material inhomogeneity $\mathbb{T}_{i\alpha\beta}(\theta^\alpha)$, which encodes the integrability of the surface material uniformity bases $\boldsymbol{a}_\alpha(\theta^\alpha)$, should be interpreted as a density of dislocations smeared over the base manifold $\omega$. On the other hand, the material inhomogeneity $\mathbb{T}_{i\alpha 3}(\theta^\alpha)$ should be interpreted as an `apparent' disclination density, smeared over the base manifold $\omega$, for they encode the compatibility of the out-of-surface material uniformity bases $\boldsymbol{d}(\theta^\alpha)$ and $\boldsymbol{D}_\alpha(\theta^\alpha)$ in the sense that a certain derivative of the transverse director field $\boldsymbol{d}(\theta^\alpha)$, defined as $\nabla_\alpha\boldsymbol{d}:= \boldsymbol{d}_{,\alpha}-\boldsymbol{D}_{\alpha}$, vanishes if and only if $\mathbb{T}_{i\alpha 3}(\theta^\alpha)=0$.\footnote{ We interpret the quantity $\nabla_\alpha\boldsymbol{d}(\theta^\alpha)$ as an `apparent' disclination density because this quantity contributes nothing to the curvature $R^i_{jkl}$ of the material space of the actual 3-d shell (while it does contribute to the curvature $K^i_{jkl}$ of its Riemannian space, hence producing incompatibility in the strain field). Only in the Cosserat picture, which is a mathematical artifact in the present case, a part of the actual 3-d dislocation density `appears as' a disclination density.} The differential operator $\nabla_\alpha$ gives rise to a parallelism, hence an Ehresmann connection \cite{eps2}, on the normal sub-bundle of the original vector bundle in the Cosserat picture. We can summarize the above result as
\begin{prop}
Under the assumptions made in the present remark, the shell is materially homogeneous if and only if $\mathbb{T}_{i\alpha j}(\theta^\alpha)$  vanish simultaneously.
\end{prop}}
  \end{rem}
  
 \begin{rem}
{\rm Our framework generalizes the earlier notion of material homogeneity, and the associated measures, in a thin shell as was proposed in \cite{eps2,edeleon}. The notion of material homogeneity discussed therein is the existence of a globally flat diffeomorphic configuration $\hat \omega$, carrying everywhere a normal director attached to $\hat \omega$. As is clearly evident from the statements of Theorems \ref{thm:1} and \ref{thm:2}, and the ensuing discussion, this should not necessarily be the general case because the diffeomorphic image of the current configuration of the shell (or, in the Cosserat picture, the image of the current vector bundle under a principal bundle isomorphism), upon stress relaxation, may not be globally flat at all.}
 \end{rem}

\noindent\textbf{Proof of Theorem 3.1:} $(i)$ According to Theorem \ref{thm:1}, hypothesis \eqref{hyp:3} implies that there exist vector fields $\boldsymbol{g}_j(\theta^\alpha,\zeta):=\hat H_{ij}\boldsymbol{G}^i$, where $[\hat H_{ij}(\theta^\alpha,\zeta)]$ is an invertible matrix field on $\mathcal{B}$, satisfying \eqref{form:1}. Let $\boldsymbol{g}_i(\theta^\alpha,\zeta)$ be analytic in $\zeta$. Then there exist vector fields on $\omega$, $\boldsymbol{a}_\alpha(\theta^\alpha):= \boldsymbol{g}_\alpha(\theta^\alpha,0)$, $\boldsymbol{D}_\alpha(\theta^\alpha):= \boldsymbol{g}_\alpha'(\theta^\alpha,0)$, $\boldsymbol{d}(\theta^\alpha):= \boldsymbol{g}_3(\theta^\alpha,0)$, and $\boldsymbol{E}_\alpha(\theta^\alpha):= \boldsymbol{g}_3'(\theta^\alpha,0)$ (superscript prime denotes a derivative with respect to $\zeta$), such that 
 \[\boldsymbol{g}_\alpha(\theta^\alpha,\zeta)=\boldsymbol{a}_\alpha(\theta^\alpha)+\zeta \boldsymbol{D}_\alpha(\theta^\alpha)+o(\zeta)~\text{and}\]
 \[\boldsymbol{g}_3(\theta^\alpha,\zeta)=\boldsymbol{d}(\theta^\alpha)+\zeta \boldsymbol{E}(\theta^\alpha)+o(\zeta).\]
 Hypothesis \eqref{hyp:4} subsequently implies that $g_{ij}(\theta^\alpha,\zeta)=\boldsymbol{g}_i(\theta^\alpha,\zeta)\cdot\boldsymbol{g}_j(\theta^\alpha,\zeta)$. From $g_{\alpha\beta}(\theta^\alpha,\zeta)=\boldsymbol{g}_\alpha(\theta^\alpha,\zeta)\cdot\boldsymbol{g}_\beta(\theta^\alpha,\zeta)$, we obtain
\begin{eqnarray}
a_{\alpha\beta}(\theta^\alpha)+\zeta P_{\alpha\beta}(\theta^\alpha)+\zeta^2 Q_{\alpha\beta}(\theta^\alpha)&=&\boldsymbol{a}_{\alpha}(\theta^\alpha)\cdot\boldsymbol{a}_{\beta}(\theta^\alpha)\nonumber\\
&&+\zeta \bigg(\boldsymbol{a}_{\alpha}(\theta^\alpha)\cdot\boldsymbol{D}_{\beta}(\theta^\alpha)+\boldsymbol{a}_{\beta}(\theta^\alpha)\cdot\boldsymbol{D}_{\alpha}(\theta^\alpha)\bigg)\nonumber\\
&&+\zeta^2 \, \boldsymbol{D}_{\alpha}(\theta^\alpha)\cdot \boldsymbol{D}_{\beta}(\theta^\alpha) + O(\zeta^3),
\label{comparison:1}
\end{eqnarray}
 which implies that $\boldsymbol{g}_\alpha(\theta^\alpha,\zeta)$ is linear in $\zeta$:
 \[\boldsymbol{g}_\alpha(\theta^\alpha,\zeta)=\boldsymbol{a}_\alpha(\theta^\alpha)+\zeta \boldsymbol{D}_\alpha(\theta^\alpha).\]
 From $g_{\alpha 3}(\theta^\alpha,\zeta)=\boldsymbol{g}_\alpha(\theta^\alpha,\zeta)\cdot\boldsymbol{g}_3(\theta^\alpha,\zeta)$, we obtain
 \begin{eqnarray}
\Delta_\alpha+\zeta\{a^{\sigma\gamma}\,\Delta_\sigma (\Lambda_{\gamma\alpha}-B_{\gamma\alpha})+\Lambda_\alpha (\Delta+1)\}&=&\boldsymbol{a}_{\alpha}(\theta^\alpha)\cdot\boldsymbol{d}(\theta^\alpha)\nonumber\\
&&+\zeta\bigg(\boldsymbol{D}_{\alpha}(\theta^\alpha)\cdot\boldsymbol{d}(\theta^\alpha)+\boldsymbol{a}_\alpha(\theta^\alpha)\cdot\boldsymbol{E}(\theta^\alpha)\bigg)\nonumber\\
&&+O(\zeta^2),
\label{comparison:2}
\end{eqnarray}
 which implies, firstly, that $g_{\alpha 3}(\theta^\alpha,\zeta)$ is linear in $\zeta$:
 \[\boldsymbol{g}_3(\theta^\alpha,\zeta)=\boldsymbol{d}(\theta^\alpha)+\zeta \boldsymbol{E}(\theta^\alpha).\]
 But, since $g_{3 3}(\theta^\alpha,\zeta)=\boldsymbol{g}_3(\theta^\alpha,\zeta)\cdot\boldsymbol{g}_3(\theta^\alpha,\zeta)$, i.e.
 \begin{equation}
  a^{\alpha\beta}\,\Delta_\alpha \Delta_\beta+(\Delta+1)^2=\boldsymbol{d}(\theta^\alpha)\cdot\boldsymbol{d}(\theta^\alpha)+O(\zeta),
  \label{comparison:3}
 \end{equation}
we obtain $\boldsymbol{E}(\theta^\alpha)=\boldsymbol{0}$ or that $\boldsymbol{g}_3(\theta^\alpha,\zeta)$ is independent of $\zeta$:
\[\boldsymbol{g}_3(\theta^\alpha,\zeta)=\boldsymbol{d}(\theta^\alpha).\]
 Finally, relations \eqref{fields} can be obtained using \eqref{comparison:1}, \eqref{comparison:2}, and \eqref{comparison:3}.

$(ii)$ We can write various components of the torsion tensor, in terms of $[\hat H_{ij}]$ and its inverse $[\hat H^{-1}]^{ij}$, as
\begin{subequations}
\begin{align}
T^\mu_{\alpha\beta}&=(\hat H^{-1})^{\mu\sigma}\hat H_{\sigma[\alpha,\beta]}+(\hat H^{-1})^{\mu 3}\hat H_{3[\alpha,\beta]},\\
T^3_{\alpha\beta}&=(\hat H^{-1})^{3\sigma}\hat H_{\sigma[\alpha,\beta]}+(\hat H^{-1})^{33}\hat H_{3[\alpha,\beta]},\\
T^\mu_{\alpha 3}&=(\hat H^{-1})^{\mu\sigma}\hat H_{\sigma[\alpha,3]}+(\hat H^{-1})^{\mu 3}\hat H_{3[\alpha,3]},~\text{and}\\
T^3_{\alpha 3}&=(\hat H^{-1})^{3\sigma}\hat H_{\sigma[\alpha,3]}+(\hat H^{-1})^{33}\hat H_{3[\alpha,3]}.
\end{align}
\end{subequations}
If $T^p_{\alpha\beta}(\theta^\alpha,0)=0$, i.e. $\hat H_{i[\alpha;\beta]}\big|_{\zeta=0}=0$ (the covariant curl will then be same as the ordinary curl), we have $\boldsymbol{g}_{\beta,\alpha}(\theta^\alpha,0)=\boldsymbol{g}_{\alpha,\beta}(\theta^\alpha,0)$, which implies that there exist three sufficiently smooth real functions $r_i(\theta^\alpha)$ such that $\hat H_{i\alpha}(\theta^\alpha,0)=r_{i,\alpha}(\theta^\alpha)$ and
\[\boldsymbol{g}_{\alpha}(\theta^\alpha,0)=\boldsymbol{a}_{\alpha}(\theta^\alpha)=\boldsymbol{r}_{,\alpha}(\theta^\alpha),\]
where $\boldsymbol{r} = r_\alpha\boldsymbol{A}^\alpha+r_3\boldsymbol{N}$.
 
Moreover, if $T^p_{\alpha 3}(\theta^\alpha,0)=0$, i.e. $\hat H_{i[\alpha;3]}\big|_{\zeta=0}=0$, we have $\boldsymbol{d}_{,\alpha}(\theta^\alpha)=\boldsymbol{g}_{\alpha,3}(\theta^\alpha,0)$ or equivalently
\[\boldsymbol{d}_{,\alpha}(\theta^\alpha)=\boldsymbol{D}_{\alpha}(\theta^\alpha).\]
 Evidently, we have the existence of a sufficiently smooth diffeomorphism $\boldsymbol{\chi}(\theta^\alpha,\zeta)=\boldsymbol{r}(\theta^\alpha)+\zeta\boldsymbol{d}(\theta^\alpha)$ on $\mathcal{B}$ such that $\boldsymbol{g}_i(\theta^\alpha,\zeta)=\boldsymbol{\chi}_{,i}(\theta^\alpha,\zeta)$.\hfill$\square$

\subsection{Strain compatibility}
For a materially homogeneous thin shell, the components of the Riemann-Christoffel curvature  $K^i_{jkl}$ of its Riemannian space $(\mathcal{B};\,g_{ij})$ vanish giving rise to the strain compatibility conditions. In this subsection, we will show that the strain compatibility relations for the 2-d strain measures can be recovered from the vanishing of $K_{ijkl}$ at $\omega$ alone, not necessarily on the full $\mathcal{B}$. A more restricted version of the following theorem has appeared in the proof of Theorem 2.8-1 in \cite{ciar1}.
\begin{thm}\label{thm:3}
 If the restriction of the Riemannian curvature of the shell to $\zeta=0$ is zero, i.e. if $K^i_{jkl}(\theta^\alpha,0)=0$, then there exists an open  sufficiently smooth diffeomorphic image $\hat{\mathcal{B}}$ of $\mathcal{B}$, parametrized by $\boldsymbol{\chi}(\theta^\alpha,\zeta)=\boldsymbol{r}(\theta^\alpha)+\zeta \boldsymbol{d}(\theta^\alpha)$, such that $\boldsymbol{g}_{i}(\theta^\alpha,\zeta)=\boldsymbol{\chi}_{,i}(\theta^\alpha,\zeta)$.
\end{thm}

\begin{rem}{\rm
(2-d strain compatibility conditions) For a materially homogeneous thin elastic shell, the intrinsic strain measures are given as
\begin{subequations}
\begin{align}
E_{\alpha\beta} &=\frac{1}{2}(a_{\alpha\beta}-A_{\alpha\beta}), \label{def1}\\
\Delta_{\alpha} &= d_\alpha,\label{def2}\\
\Delta &= d-1,\label{def3}\\
\Lambda_{\alpha\beta} &= B_{\alpha\beta}+d_{\alpha|\beta}-d\,b_{\alpha\beta},~\text{and}\label{def4}\\
\Lambda_{\alpha} &= d_{,\alpha}+d_\sigma\,b^\sigma_\alpha,\label{def5}
 \end{align}
 \label{def}%
 \end{subequations}
where $\boldsymbol{d}=d_\alpha\boldsymbol{a}^\alpha+d\boldsymbol{n}$ and $a_{\alpha\beta}=\boldsymbol{r}_{,\alpha}\cdot\boldsymbol{r}_{,\beta}$. Viewed as a set of partial differential equations for $\boldsymbol{r}(\theta^\alpha)$ and $\boldsymbol{d}(\theta^\alpha)$, which together represent a directed surface $\hat{\omega}$, the integrability conditions of the system \eqref{def} give us the compatibility conditions for the 2-d strain measures $E_{\alpha\beta}$, $\Lambda_{\alpha\beta}$, $\Lambda_\alpha$, $\Delta_\alpha$ and $\Delta$ of a sufficiently thin shell.  We now discuss these conditions, following Epstein \cite{eps1}, before moving on to the proof of the theorem. The question of compatibility can be posed as: given sufficiently smooth fields including a symmetric matrix $[E_{\alpha\beta}]$, an invertible matrix $[\Lambda_{\alpha\beta}]$, two vectors $\{\Delta_\alpha\}$ and $\{\Lambda_\alpha\}$, and a scalar $\Delta (\ne -1)$ on a parametrized surface $\omega$, with its first and second fundamental form as $A_{\alpha\beta}(\theta^\alpha)$ and $B_{\alpha\beta}(\theta^\alpha)$, respectively, what are the conditions to be satisfied by the five given fields for the existence of a sufficiently smooth parametrized surface $\hat{\omega}$, along with a director field $\boldsymbol{d}(\theta^\alpha)$, having its first fundamental form given by $a_{\alpha\beta}=A_{\alpha\beta}+2 E_{\alpha\beta}$ and its second fundamental form $b_{\alpha\beta}$ suitably constructed out of the given fields, so that the equations \eqref{def} are satisfied. Towards this end, let us first ensure the existence of surface $\hat{\omega}$. We can read off the formula for its second fundamental form $b_{\alpha\beta}$ from the definitions \eqref{def2}, \eqref{def2},  and \eqref{def3} as
\begin{equation}
b_{\alpha\beta}=-\frac{\Lambda_{\alpha\beta}-B_{\alpha\beta}-\Delta_{\alpha|\beta}}{\Delta+1}.\label{sff}
\end{equation}
To ensure that the given fields indeed give rise, via \eqref{sff}, to an admissible  second fundamental form for a realizable surface, we must ensure that $b_{[\alpha\beta]}=0$, or equivalently
\begin{equation}
J:= e^{\alpha\beta}b_{\alpha\beta}=0\,\,\,\textrm{or}\,\,\,\Lambda_{[\alpha\beta]}-\Delta_{[\alpha|\beta]}=0,\label{comp1}
\end{equation}
where $e^{\alpha\beta}=e_{\alpha\beta}$ is the 2-d permutation symbol. Equation \eqref{comp1}$_2$ is the first strain compatibility condition. Keeping \eqref{comp1} in mind, the conditions on $a_{\alpha\beta}=A_{\alpha\beta}+2E_{\alpha\beta}$ and $b_{(\alpha\beta)}$ so that they indeed constitute the first and the second fundamental forms of a parametrized surface $\hat{\omega}$, upto isometries of $\mathbb{R}^3$, are the Gauss and Codazzi-Mainardi relations:
\begin{equation}
 S_{\tau\alpha\beta\sigma}=b_{(\alpha\sigma)}b_{(\beta\tau)}-b_{(\alpha\beta)}b_{(\sigma\tau)}~\text{and}\label{comp2}
\end{equation}
\begin{equation}
 b_{(\alpha\sigma),\beta}-b_{(\alpha\beta),\sigma}+s^\mu_{\alpha\sigma}b_{(\beta\mu)}-s^\mu_{\alpha\beta}b_{(\sigma\mu)}=0,\label{comp3}
\end{equation}
provided $\omega$ is simply connected, where
\[S_{\tau\alpha\beta\sigma}:= s_{\alpha\sigma\tau,\beta}-s_{\alpha\beta\tau,\sigma}+s^\mu_{\alpha\beta}s_{\sigma\tau\mu}-s^\mu_{\alpha\sigma}s_{\beta\tau\mu},\]
\[s_{\alpha\beta\mu}:= \frac{1}{2}(a_{\alpha\mu,\beta}+a_{\beta\mu,\alpha}-a_{\alpha\beta,\mu}),\text{and}\]
\[s^\sigma_{\alpha\beta}:= a^{\sigma\mu}s_{\alpha\beta\mu}.\]
Equations \eqref{comp2} and \eqref{comp3} are the second and the third strain compatibility conditions. The three compatibility conditions derived so far ensure existence of a unique (modulo an isometry in $\mathbb{R}^3$) surface $\hat{\omega}$.
Finally, to ensure the existence of a director field $\boldsymbol{d}(\theta^\alpha)$ on $\hat{\omega}$, consistent with \eqref{def}, we require
\begin{equation}
I_\beta:=\Lambda_\beta-\Delta_{,\beta}+\Delta_\alpha a^{\alpha\gamma}\bigg(\frac{\Lambda_{(\gamma\beta)}-B_{\gamma\beta}-\Delta_{(\gamma|\beta)}}{\Delta+1}\bigg)=0.
\label{comp4}
\end{equation}
This is the fourth compatibility condition, obtained by eliminating $b_{(\alpha\beta)}$ (recall that $b_{[\alpha\beta]}=0$) from equations \eqref{def2}-\eqref{def5}. Altogether, we have six independent strain compatibility relations for twelve independent components of strain measures.}

\end{rem}

\begin{rem}{\rm
(Residual stress field)
Given the strain energy density $W(E_{ij})$ per unit volume of the shell material, the 2-d strain energy density per unit area on $\omega$, denoted by $\psi(E_{\alpha\beta},\Lambda_{\alpha\beta},\Lambda_\alpha,\Delta_\alpha,\Delta)$ (or equivalently as $U(\boldsymbol{a}_\alpha, \boldsymbol{D}_\alpha, \boldsymbol{d})$),  can be calculated via the following integration \cite{greennaghdi}
 \begin{equation}
 \psi(E_{\alpha\beta},\Lambda_{\alpha\beta},\Lambda_\alpha,\Delta_\alpha,\Delta)=\frac{1}{\sqrt{A}}\int^h_{-h}\sqrt{G}W(E_{ij})\,d\zeta.
 \end{equation}
Then, with zero body force, the 2-d equilibrium equations for various `stress' measures of the shell are
  \begin{subequations}
  \begin{align}
   \boldsymbol{T}^\alpha_{;\alpha}&=\boldsymbol{0},\\
   \boldsymbol{M}^\alpha_{;\alpha}-\boldsymbol{k}&=\boldsymbol{0},~\text{and}\\
   \boldsymbol{a}_\alpha\times\boldsymbol{T}^\alpha+\boldsymbol{D}_\alpha\times\boldsymbol{M}^\alpha+\boldsymbol{d}\times\boldsymbol{k}&=\boldsymbol{0},
  \end{align}
\label{shell:eqb}%
\end{subequations}
where $j\boldsymbol{T}^\alpha=\partial_{\boldsymbol{a}_\alpha} U$, $j\boldsymbol{M}^\alpha=\partial_{\boldsymbol{D}_\alpha} U$, $j\boldsymbol{k}=\partial_{\boldsymbol{d}} U$, $j := \sqrt{\frac{a}{A}}$, $a:=det(a_{\alpha\beta})$, $A:=det(A_{\alpha\beta})$, and $G:=det(G_{ij})$. If in addition the shell is materially inhomogeneous then the compatibility relations are not satisfied and provide additional equations for determination of the stresses, provided the incompatibility is known. We introduce six incompatibility measures $\mathbb{J}$, $\mathbb{K}$, $\mathbb{L}_\sigma$, and $\mathbb{I}_{\alpha}$ such that the strain compatibility equations take the form
\begin{subequations}
\begin{align}
J&=\mathbb{J}(\theta^\alpha),\\
 S_{1212}-b_{11}b_{22}+b_{(12)}^2&=\mathbb{K}(\theta^\alpha),\\
  b_{(\sigma 1)|2}-b_{(\sigma 2)|1} &=\mathbb{L}_\sigma(\theta^\alpha),~\text{and}\\
  I_\alpha&=\mathbb{I}_{\alpha}(\theta^\alpha).
\end{align}
\label{shell:comp}%
\end{subequations}
The incompatibilities are related to curvature fields $K_{1212}$, $K_{12\sigma3}$, and $K_{\rho3\sigma3}$ according to the following six relations:
\begin{equation}
\mathbb{K}(\theta^\alpha) = K_{1212}(\theta^\alpha,0),
\end{equation}
\begin{equation}
a^{2\beta}\Delta_\beta\mathbb{K}+(\Delta+1)\mathbb{L}_1=K_{1213}(\theta^\alpha,0),
\end{equation}
\begin{equation}
-a^{1\beta}\Delta_\beta\mathbb{K}+(\Delta+1)\mathbb{L}_2= K_{1223}(\theta^\alpha,0),~\text{and}
\end{equation}
\begin{eqnarray}
 &&(\Delta+1)\, \mathbb{I}_{(\rho|\sigma)}-\Lambda_{(\rho}\mathbb{I}_{\sigma)}-a^{\alpha\beta}\Delta_\alpha\bigg\{ b_{(\beta\big(\rho)}\mathbb{I}_{\sigma\big)}+e_{[\beta\big(\rho]}\mathbb{I}_{\sigma\big)}\mathbb{J}- b_{(\rho\sigma)}\,\mathbb{I}_\beta\bigg\} \nonumber\\
 &&+(\Delta+1)\mathbb{J}\bigg( a^{\alpha\beta}+\frac{a^{\alpha\mu}a^{\beta\nu}\Delta_\mu \Delta_\nu}{(\Delta+1)^2}\bigg)\,\bigg((\Delta+1)\mathbb{J}\,e_{\alpha\rho}e_{\beta\sigma}\nonumber\\
 &&+e_{\alpha\rho}(\Lambda_{\beta\sigma}-B_{\beta\sigma})+e_{\beta\sigma}(\Lambda_{\alpha\rho}-B_{\alpha\rho})\bigg)=K_{\rho3\sigma3}(\theta^\alpha,0).
\end{eqnarray}
The above construction is analogous to Kr\"{o}ner's framework \cite{kroner81} of residual stress determination for a given incompatibility tensor field in the context of 3-d linear elasticity. The incompatibilities can also be obtained from defect densities. For instance, consider the case when only dislocation anomalies are present in the 3-d shell. Then incompatibilities can be written in terms of the torsion tensor as
\begin{equation}
\mathbb{K}(\theta^\alpha) = -g_{p1}(\theta^\alpha,0)[C^p_{22|1}-C^p_{21|2}+C^h_{22}C^p_{h1}-C^h_{21}C^p_{h2}]\big|_{\zeta=0},
\end{equation}
\begin{equation}
a^{2\beta}\Delta_\beta\mathbb{K}+(\Delta+1)\mathbb{L}_1=-g_{p1}(\theta^\alpha,0)[C^p_{23|1}-C^p_{21|3}+C^h_{23}C^p_{h1}-C^h_{21}C^p_{h3}]\big|_{\zeta=0},
\end{equation}
\begin{equation}
-a^{1\beta}\Delta_\beta\mathbb{K}+(\Delta+1)\mathbb{L}_2= -g_{p1}(\theta^\alpha,0)[C^p_{23|2}-C^p_{22|3}+C^h_{23}C^p_{h2}-C^h_{22}C^p_{h3}]\big|_{\zeta=0},~\text{and}
\end{equation}
\begin{eqnarray}
 &&(\Delta+1)\, \mathbb{I}_{(\rho|\sigma)}-\Lambda_{(\rho}\mathbb{I}_{\sigma)}-a^{\alpha\beta}\Delta_\alpha\bigg\{ b_{(\beta\big(\rho)}\mathbb{I}_{\sigma\big)}+e_{[\beta\big(\rho]}\mathbb{I}_{\sigma\big)}\mathbb{J}- b_{(\rho\sigma)}\,\mathbb{I}_\beta\bigg\} \nonumber\\
 &&+(\Delta+1)\mathbb{J}\bigg( a^{\alpha\beta}+\frac{a^{\alpha\mu}a^{\beta\nu}\Delta_\mu \Delta_\nu}{(\Delta+1)^2}\bigg)\,\bigg((\Delta+1)\mathbb{J}\,e_{\alpha\rho}e_{\beta\sigma}\nonumber\\
 &&+e_{\alpha\rho}(\Lambda_{\beta\sigma}-B_{\beta\sigma})+e_{\beta\sigma}(\Lambda_{\alpha\rho}-B_{\alpha\rho})\bigg)\nonumber\\
 &=&-g_{p\rho}(\theta^\alpha,0)[C^p_{33|\sigma}-C^p_{3\sigma|3}+C^h_{33}C^p_{h\sigma}-C^h_{3\sigma}C^p_{h3}]\big|_{\zeta=0}. \label{incomptor4}
\end{eqnarray}

}

\end{rem}


\noindent\textbf{Proof of Theorem 3.2:} The coefficients of the Levi-Civita connection, defined by $\Gamma_{ijp}:=\frac{1}{2}(g_{ip,j}+g_{jp,i}-g_{ij,p})$, are
\begin{subequations}
\begin{align}
\Gamma_{333} &=0,~\Gamma_{33\rho} =U_\rho-\frac{1}{2}V_{,\rho},~\Gamma_{3\rho 3} =\Gamma_{\rho 33}=\frac{1}{2}V_{,\rho},\label{a}\\
\Gamma_{3\rho\sigma} &=\Gamma_{\rho 3\sigma}=\Delta_{[\sigma,\rho]}+\frac{1}{2}P_{\rho\sigma}+\zeta\,\bigg(U_{[\sigma,\rho]}+Q_{\rho\sigma}\bigg),\\
\Gamma_{\rho\sigma 3} &=\Delta_{(\sigma,\rho)}-\frac{1}{2}P_{\rho\sigma}+\zeta\,\bigg(U_{(\sigma,\rho)}-Q_{\rho\sigma}\bigg),~\text{and}\\
\Gamma_{\rho\sigma \delta} &=s_{\rho\sigma\delta}+\frac{\zeta}{2}\bigg(P_{\rho\delta,\sigma}+P_{\sigma\delta,\rho}-P_{\sigma\rho,\delta}\bigg)               +\frac{\zeta^2}{2}\bigg(Q_{\rho\delta,\sigma}+Q_{\sigma\delta,\rho}-Q_{\sigma\rho,\delta}\bigg).
\end{align}
\label{levivicita}%
\end{subequations}
The curvature $K_{ijkl}$ has six independent components such that $K_{ijkl}=0$ if and only if $K_{1212}=0$, $K_{12\sigma 3}=0$, and $K_{\rho 3 \sigma 3}=0$.
After some manipulations, it can be shown that
\begin{eqnarray}
K_{1212}\big|_{\zeta=0} &:=& \big(\Gamma_{221,1}-\Gamma_{211,2}+\Gamma^i_{21}\Gamma_{21 i}-\Gamma^i_{22}\Gamma_{11 i}\big)\big|_{\zeta=0}\nonumber\\
&=&S_{1212}-[b_{11}b_{22}-b_{(12)}^2].
\end{eqnarray}
Hence, $K_{1212}\big|_{\zeta=0}=0$ implies that 
\begin{equation}
 S_{1212}=b_{11}b_{22}-b_{(12)}^2,
 \label{gauss}
\end{equation}
which is the single independent Gauss' equation, cf. \eqref{comp2}.
Moreover, 
\begin{eqnarray}
K_{12\sigma 3}\big|_{\zeta=0}&:=&\big(\Gamma_{231,\sigma}-\Gamma_{2\sigma 1,3}+\Gamma^i_{2\sigma}\Gamma_{31 i}-\Gamma^i_{23}\Gamma_{\sigma1 i}\big)\big|_{\zeta=0}\nonumber\\
&=&a^{\alpha\beta}\Delta_\beta\bigg(S_{\alpha\sigma 21}+\big[b_{(\sigma 2)}b_{(\alpha 1)}-b_{(\sigma 1)} b_{(\alpha 2)}\big]\bigg)+(\Delta+1)\big[b_{(\sigma 1)|2}-b_{(\sigma 2)|1}\big],
\end{eqnarray}
which can be used to calculate
\begin{equation}
K_{121 3}\big|_{\zeta=0}=a^{2\beta}\Delta_\beta\bigg(S_{2 1 21}+\big[b_{(1 2)}b_{(2 1)}-b_{(1 1)} b_{(2 2)}\big]\bigg)+(\Delta+1)\big[b_{(1 1)|2}-b_{(1 2)|1}\big]~\text{and}
\label{rel1}
\end{equation}
\begin{equation}
K_{122 3}\big|_{\zeta=0}=-a^{1\beta}\Delta_\beta\bigg(S_{1 2 12}+\big[b_{(21)}b_{(1 2)}-b_{(1 1)} b_{(2 2)}\big]\bigg)+(\Delta+1)\big[b_{(2 1)|2}-b_{(2 2)|1}\big].
\label{rel2}
\end{equation}
Substituting \eqref{gauss} in \eqref{rel1} and \eqref{rel2}, along with $K_{12\sigma 3}\big|_{\zeta=0}=0$, yield
\begin{equation}
 b_{(1 1)|2}-b_{(1 2)|1}=0\,\,\,\textrm{and}\,\,\,\,b_{(2 1)|2}-b_{(2 2)|1}=0,
 \label{codmain}
\end{equation}
which are two independent Codazzi-Mainardi equations, cf. \eqref{comp3}.
Finally, we consider
\begin{eqnarray}
 K_{\rho 3 \sigma 3}\big|_{\zeta=0}&:=& \big(\Gamma_{33\rho,\sigma}-\Gamma_{3\sigma\rho,3}+\Gamma^i_{3\sigma}\Gamma_{3\rho i}-\Gamma^i_{33}\Gamma_{\sigma\rho i}\big)\big|_{\zeta=0}\nonumber\\
 &=&(\Delta+1)\, I_{(\rho|\sigma)}-\Lambda_{(\rho}I_{\sigma)}-a^{\alpha\beta}\Delta_\alpha\bigg\{b_{(\beta\big(\rho)}I_{\sigma\big)}+\frac{1}{2}e_{\beta\big(\rho}I_{\sigma\big)}J- b_{(\rho\sigma)}\,I_\beta\bigg\} \nonumber\\
 &&+\bigg( a^{\alpha\beta}+\frac{a^{\alpha\mu}a^{\beta\nu}\Delta_\mu \Delta_\nu}{(\Delta+1)^2}\bigg)\,(\Delta+1)J\bigg((\Delta+1)Je_{\alpha\rho}e_{\beta\sigma}\nonumber\\
 &&+e_{\alpha\rho}(\Lambda_{\beta\sigma}-B_{\beta\sigma})+b_{\beta\sigma}(\Lambda_{\alpha\rho}-B_{\alpha\rho})\bigg).
 \label{rel3}
\end{eqnarray}
With $K_{\rho 3 \sigma 3}\big|_{\zeta=0}=0$, \eqref{rel3} is a set of three coupled first order homogeneous non-linear partial differential algebraic equations (PDAE) for three unknowns $I_\alpha$ and $J$. Note that

$(a)$ $I_\alpha=0$ and $J=0$ are solutions to this PDAE, but there can be other non-zero solutions whose nature depends strongly on the coefficient functions.

$(b)$ If we assume $J=0$, the PDAE reduce down to a first order homogeneous overdetermined system of linear PDEs in $I_\alpha$. The system has a zero solution and other non-zero solutions depending on the coefficient functions. We disregard the unphysical non-zero solutions because they become unbounded under generic perturbations of the initial condition for generic coefficient functions \cite{bryant}.

$(c)$ If we assume $I_\alpha=0$, the PDAE reduce down to three quadratic algebraic equations in $J$ which can be easily shown to have the unique solution $J=0$.

These three facts imply that $I_\alpha$ must be proportional to $J$, i.e. there exist functions $\mathcal{L}_\alpha(J;\theta^\alpha)$ such that $I_{\alpha}=\mathcal{L}_\alpha(J;\theta^\alpha)J$. Using this in the PDAE reduces it to a set of overdetermined first order non-linear PDEs in $J$ which clearly has a zero solution along with other unphysical non-zero solutions. The zero solution implies $I_\alpha=0$.\hfill$\square$


\section{Kirchhoff-Love shell with continuous distribution of dislocations}
We now restrict ourselves to the case when the inhomogeneous shell has only dislocation anomalies and the stress relaxation process respects the Kirchhoff-Love constraint $\boldsymbol{d}=\boldsymbol{n}$, i.e. $\Delta_\alpha=\Delta=\Lambda_\alpha=0$. As a result, the Cosserat material uniformity basis $\boldsymbol{a}_\alpha$ determines the complete Cosserat material uniformity bases. The compatibility relations $I_\alpha = 0$ are trivially satisfied; hence, $\mathbb{I}_\alpha$ cannot take a non-zero value. In addition, the components $T^i_{\alpha3}$ (or equivalently $\mathbb{T}_{i3\alpha}$) of the torsion tensor are identically zero as the director fields are compatible in the sense that $\nabla_\alpha\boldsymbol{n}\equiv \boldsymbol{0}$. Under the present simplification, \eqref{incomptor4} reduces to 
\begin{equation}
a^{\alpha\beta}\mathbb{J}\,\bigg(\mathbb{J}\,e_{\alpha\rho}e_{\beta\sigma}
 +e_{\alpha\rho}(\Lambda_{\beta\sigma}-B_{\beta\sigma})+e_{\beta\sigma}(\Lambda_{\alpha\rho}-B_{\alpha\rho})\bigg)
 =0.
 \end{equation}
This relation implies that either $a^{\alpha\beta}\,\big(\mathbb{J}\,e_{\alpha\rho}e_{\beta\sigma}
 +e_{\alpha\rho}(\Lambda_{\beta\sigma}-B_{\beta\sigma})+e_{\beta\sigma}(\Lambda_{\alpha\rho}-B_{\alpha\rho})\big)=0$, a system of three equations which cannot be solved for $\mathbb{J}$ (and hence to be discarded), or $\mathbb{J}=0$, i.e. $b_{12}=b_{21}$. Hence, for a Kirchhoff-Love shell, $b_{\alpha\beta}$ is necessarily symmetric (or in other words $\Lambda_{\alpha\beta}=\Lambda_{\beta\alpha}$). The three non-trivial strain incompatibility relations are
\begin{subequations}
\begin{align}
 S_{1212}-b_{11}b_{22}+b_{12}^2&=K_{1212}(\theta^\alpha,0)~\text{and}\\
  b_{\sigma 1|2}-b_{\sigma 2|1} &=K_{12\sigma3}(\theta^\alpha,0).
\end{align}
\label{shell:comp:Kirchhoff-love:iso}%
\end{subequations}
To rewrite the right hand side of the above relations in terms of dislocation density (torsion) we note that presently
\begin{equation}
C^\rho_{\alpha\beta}(\theta^\alpha,\zeta)= T^\rho_{\alpha \beta}(\theta^\alpha,0)-g^{\mu\rho}(\theta^\alpha,\zeta)\,g_{\alpha \nu}(\theta^\alpha,\zeta)\,T^\nu_{\mu\beta}(\theta^\alpha,0)-g^{\mu\rho}(\theta^\alpha,\zeta)\,g_{\beta \nu}(\theta^\alpha,\zeta)\,T^\nu_{\mu\alpha}(\theta^\alpha,0)
\end{equation}
\begin{equation}
~\text{and}~C^3_{\alpha\beta}(\theta^\alpha,\zeta)= T^3_{\alpha \beta}(\theta^\alpha,0)
\end{equation}
are the only non-zero components of the contortion tensor. Also, 
\begin{equation}
 g_{\alpha\beta}(\theta^\alpha,\zeta)=a_{\alpha\beta}(\theta^\alpha)-2\zeta\,b_{\alpha\beta}(\theta^\alpha)+o(\zeta)~\text{and}~g_{i3}(\theta^\alpha,\zeta)=\delta_{i3},
\end{equation}
therefore,
\begin{equation}
 g^{\alpha\beta}(\theta^\alpha,\zeta)=a^{\alpha\beta}(\theta^\alpha)+2\zeta\,a^{\alpha\sigma}a^{\gamma\beta}b_{\sigma\gamma}(\theta^\alpha)+o(\zeta)~\text{and}~g^{i3}(\theta^\alpha,\zeta)=\delta_{i3}.
\end{equation}
The strain incompatibility relations \eqref{shell:comp:Kirchhoff-love:iso} can be rewritten as
\begin{subequations}
\begin{align}
 S_{1212}-b_{11}b_{22}+b_{12}^2&=-a_{\rho 1}(\theta^\alpha)[C^\rho_{22|1}-C^\rho_{21|2}+C^\mu_{22}C^\rho_{\mu1}-C^\mu_{21}C^\rho_{\mu2}]\big|_{\zeta=0}\\
  \text{and}~b_{(\sigma 1)|2}-b_{(\sigma 2)|1} &=a_{\rho1}(\theta^\alpha)C^\rho_{2\sigma|3}(\theta^\alpha,0).
\end{align}
\label{shell:comp:Kirchhoff-love:aniso}%
\end{subequations}

With zero body force distribution and negligible inertia, the equilibrium equations for a Kirchhoff-Love shell take the form \cite{steigmann14}:
\begin{subequations}
 \begin{align}
  (\sigma^{\mu\alpha}+M^{\beta\alpha}b^\mu_\beta)_{|\alpha}+M^{\beta\alpha}_{|\beta}b^\mu_\alpha &= 0~\text{and}\\  
  (\sigma^{\beta\alpha}+M^{\mu\alpha}b^\beta_\mu)b_{\beta\alpha}-M^{\beta\alpha}_{|\beta\alpha} &= 0,
 \end{align}
 \label{eqb:Kirchhoff-love}%
\end{subequations}
where
\begin{equation}
 j\sigma^{\beta\alpha}=\frac{1}{2}\bigg(\frac{\partial\psi}{\partial E_{\alpha\beta}}+\frac{\partial\psi}{\partial E_{\beta\alpha}}\bigg)\,\,\,\textrm{and}\,\,\,
jM^{\beta\alpha}=-\frac{1}{2}\bigg(\frac{\partial\psi}{\partial \Lambda_{\alpha\beta}}+\frac{\partial\psi}{\partial \Lambda_{\beta\alpha}}\bigg).
\end{equation}
Equations \eqref{shell:comp:Kirchhoff-love:aniso} and \eqref{eqb:Kirchhoff-love} form the governing equations for the residual stress field $\sigma^{\alpha\beta}$ and bending moment field $M^{\alpha\beta}$ for a Kirchhoff-Love shell with a continuous distribution of surface dislocation field specified by $T^\rho_{\alpha\beta}(\theta^\alpha,0)$.
Dimensional analysis and representation theorem show that  for a sufficiently thin isotropic Kirchhoff-Love shell, $\psi(E_{\alpha\beta},\Lambda_{\alpha\beta})$ can be expressed as \cite{steigmann99}
\begin{equation}
 \psi(E_{\alpha\beta},\Lambda_{\alpha\beta})=Eh\bigg(C(J_1,J_2)+h^2\sum^7_{i=3}J_i D_i(J_1,J_2)\bigg),
\end{equation}
where $E$ is the Young's modulus of the shell material and $J_1:=E_{\alpha\beta}A^{\alpha\beta}$, $J_2:=E_{\alpha\beta}E_{\mu\nu}A^{\alpha\mu}A^{\beta\nu}$, $J_3:=(\Lambda_{\alpha\beta}A^{\alpha\beta})^2$, $J_4:=\Lambda_{\alpha\beta}\Lambda_{\mu\nu}A^{\alpha\mu}A^{\beta\nu}$, $J_5:=(E_{\alpha\beta}\Lambda_{\mu\nu}A^{\alpha\mu}A^{\beta\nu})^2$, $J_6:=A^{-1}(e^{\alpha\gamma}\Lambda_{\alpha\beta}E_{\mu\nu}A_{\sigma\gamma}A^{\sigma\mu}A^{\beta\nu})^2$ and $J_7:=E_{\alpha\beta}\Lambda_{\rho\sigma}\Lambda_{\mu\nu}A^{\rho\sigma}A^{\alpha\mu}A^{\beta\nu}$. Here, $C$ and $D_i$ are dimensionless functions.

 \begin{rem}{\rm
 The components $T^3_{\alpha\beta}(\theta^\alpha,0)$ of the torsion tensor do not contribute to the elastic deformation of a conventional Kirchhoff-Love shell. In other words, a Kirchhoff-Love shell geometrically admits only in-surface dislocation density represented by $T^\rho_{\alpha\beta}(\theta^\alpha,0)$. This is also true for Kirchhoff-Love shells with uniform thickness distention, i.e. when $\boldsymbol{d}=(\Delta+1)\boldsymbol{n}$ with constant $\Delta$.}
\end{rem}

\begin{rem} {\rm (Pure bending of an isotropic Kirchhoff-Love plate)
In case of pure bending of a plate, $a_{\alpha\beta}=A_{\alpha\beta}$ and $B_{\alpha\beta}=0$. The curvilinear coordinates $(\theta^1,\theta^2)$ can be identified with Cartesian coordinates; hence, $a_{\alpha\beta}=A_{\alpha\beta}=\delta_{\alpha\beta}$ and $s_{\alpha\beta\gamma}=0$. Moreover, $S_{1212}=0$. The strain incompatibility equations \eqref{shell:comp:Kirchhoff-love:iso} are reduced to
\begin{subequations}
\begin{align}
 -\Lambda_{11}\Lambda_{22}+\Lambda_{12}^2&=K_{1212}(\theta^\alpha,0)~\text{and}\\
  -\Lambda_{\sigma 1,2}+\Lambda_{\sigma 2,1} &=K_{12\sigma3}(\theta^\alpha,0).
\end{align}
\end{subequations}
The equilibrium equations \eqref{eqb:Kirchhoff-love} become \cite{steigmann14}
\begin{subequations}
 \begin{align}
  (\bar\sigma^{\mu\alpha}+M^{\beta\alpha}b^\mu_\beta)_{,\alpha}+M^{\beta\alpha}_{,\beta}b^\mu_\alpha &= 0~\text{and}\\  
  (\bar\sigma^{\beta\alpha}+M^{\mu\alpha}b^\beta_\mu)b_{\beta\alpha}-M^{\beta\alpha}_{,\beta\alpha} &= 0,
 \end{align}
 \label{eqb:Kirchhoff-love:purebending}%
\end{subequations}
where  $\sigma^{\alpha\beta}$ are to be interpreted as Lagrange multipliers $\bar\sigma^{\alpha\beta}(\theta^\alpha)$ associated with the deformation constraint $E_{\alpha\beta}=0$; these are determined \textit{a posteriori} after solving the complete boundary value problem. The bending moments $M^{\alpha\beta}$ are constitutively determined from the energy function $\psi(\Lambda_{\alpha\beta})$:
\begin{equation}
M^{\beta\alpha}=-\frac{1}{2}\bigg(\frac{\partial\psi}{\partial \Lambda_{\alpha\beta}}+\frac{\partial\psi}{\partial \Lambda_{\beta\alpha}}\bigg),
\end{equation}
where
\begin{equation}
 \psi(\Lambda_{\alpha\beta})=\frac{Eh^3}{24(1-\nu^2)}\bigg(\nu \Lambda_{\alpha\alpha}\Lambda_{\beta\beta}+(1-\nu)\Lambda_{\alpha\beta}\Lambda_{\alpha\beta}\bigg),
\end{equation}
$\nu$ is the Poisson's ratio of the shell material.}
\end{rem}

\begin{rem}{\rm
(Small strain large rotation of an isotropic Kirchhoff-Love plate)
Let $E_{\alpha\beta}$, and its spatial derivatives upto second order, be $O(\epsilon)$, where $\epsilon:=\frac{h}{R} <<1$ and $R$ is the minimum principal radius of curvature of the shell mid-surface for a given deformation. Let $\Lambda_{\alpha\beta}$, and its spatial derivatives upto first order,  be $O(\epsilon^\frac{1}{2})$. We identify $(\theta^1,\theta^2)$ with the Cartesian coordinates on $\omega\subset\mathbb{R}^2$. Hence, $A_{\alpha\beta}=\delta_{\alpha\beta}$ and $B_{\alpha\beta}=0$. A straightforward calculation shows that, upto $O(\epsilon)$,
\begin{equation}
 s^\tau_{\alpha\beta}:=\frac{1}{2}a^{\tau\sigma}(a_{\sigma\beta,\alpha}+a_{\sigma\alpha,\beta}-a_{\alpha\beta,\sigma})\approx A^{\tau\sigma}(E_{\sigma\beta,\alpha}+E_{\sigma\alpha,\beta}-E_{\alpha\beta,\sigma}).
 \label{levicivitaorder}
\end{equation}
Consequently, the strain incompatibility equations for an isotropic Kirchhoff-Love plate take the form (upto $O(\epsilon)$)
\begin{subequations}
\begin{align}
 2E_{12,12}-E_{11,22}-E_{22,11}-\Lambda_{11}\Lambda_{22}+\Lambda_{12}^2&=K_{1212}(\theta^\alpha,0)~\text{and}\\
  -\Lambda_{\sigma 1,2}+\Lambda_{\sigma 2,1} &=K_{12\sigma3}(\theta^\alpha,0).
\end{align}
\end{subequations}
Moreover, $J_1=O(\epsilon)$, $J_2=O(\epsilon^2)$, $J_3=O(\epsilon)$, $J_4=O(\epsilon)$, $J_5=O(\epsilon^{2.25})$, $J_6=O(\epsilon^{2.25})$, and $J_7=O(\epsilon^2)$. The quadratic strain energy function, upto $O(\epsilon^2)$, neglecting the non-conventional coupling term $J_7$ (it is important to note that, unlike the small deformation theory where the strain energy is decoupled at $O(\epsilon^2)$, a coupling term is always present), is
\begin{equation}
 \psi(E_{\alpha\beta},\Lambda_{\alpha\beta})=\frac{Eh}{2(1-\nu^2)}\bigg(\nu E_{\alpha\alpha}E_{\beta\beta}+(1-\nu)E_{\alpha\beta}E_{\alpha\beta}\bigg)+\frac{Eh^3}{24(1-\nu^2)}\bigg(\nu \Lambda_{\alpha\alpha}\Lambda_{\beta\beta}+(1-\nu)\Lambda_{\alpha\beta}\Lambda_{\alpha\beta}\bigg).
\end{equation}
The corresponding equilibrium equation in terms of $E_{\alpha\beta}$ and $\Lambda_{\alpha\beta}$ can be written upto $O(\epsilon)$ using \eqref{eqb:Kirchhoff-love}, \eqref{levicivitaorder}, and the above energy function.}
\end{rem}


\vspace{5mm}
\noindent \textbf{Acknowledgement:} We express our gratitude to Prof. Robert Bryant for the insightful discussion at mathoverflow.net \cite{bryant} about existence and uniqueness issues of the non-linear PDEs encountered in the proof of Theorem 3.2.

\bibliographystyle{plain}
\bibliography{inhomogeneity}

\begin{thebibliography}{10}

\bibitem{bilby}
B.~A. Bilby, R.~M. Bullough, and E.~Smith.
\newblock Continuous distributions of dislocations: a new application of the
  methods of non--{R}iemannian geometry.
\newblock {\em Proceedings of the Royal Society at London A}, 231:263--273,
  1955.

\bibitem{bryant}
R~Bryant.
\newblock Existence and uniqueness of a quasi-linear pde system on a surface.
\newblock MathOverflow.
\newblock URL:http://mathoverflow.net/q/198435 (version: 2015-02-28).

\bibitem{ciar1}
P~G Ciarlet.
\newblock An introduction to differential geometry with applications to
  elasticity.
\newblock {\em Journal of Elasticity}, 78-79:1--215, 2005.

\bibitem{derezin}
S~Derezin.
\newblock Gauss-{C}odazzi equations for thin films and nanotubes containing
  defects.
\newblock In H~Altenbach and V~A Eremeyev, editors, {\em Shell-like
  structures}, pages 531--547. Springer-Verlag, Berlin Heidelberg, 2011.

\bibitem{efrati}
E~Efrati, E~Sharon, and R~Kupferman.
\newblock Elastic theory of unconstrained non-{E}uclidean plates.
\newblock {\em Journal of the Mechanics and Physics of Solids}, 57:762--775,
  2009.

\bibitem{eps1}
M~Epstein.
\newblock A note on nonlinear compatibility equations for sandwich shells and
  {C}osserat surfaces.
\newblock {\em Acta Mechanica}, 31:285--289, 1979.

\bibitem{eps2}
M~Epstein and M~de~Le\'on.
\newblock Uniformity and homogeneity of elastic rods, shells and {C}osserat
  three-dimensional bodies.
\newblock {\em Archivum Mathematicum}, 32:267--280, 1996.

\bibitem{edeleon}
M~Epstein and M~de~Le\'on.
\newblock On uniformity of shells.
\newblock {\em International Journal of Solids and Structures}, 35:2173--2182,
  1998.

\bibitem{epsbook}
M~Epstein and M~Elzanowski.
\newblock {\em Material Inhomogeneities and their Evolution, A Geometric
  Approach}.
\newblock Springer-Verlag, Berlin Heidelberg, 2007.

\bibitem{epsarc1}
M~Epstein and A~Roychowdhury.
\newblock On the notion of embedded homogeneity of thin structures.
\newblock {\em Mathematics and Mechanics of Solids}, doi:
  10.1177/1081286514535127, 2014.

\bibitem{epsarc2}
M~Epstein and A~Roychowdhury.
\newblock Embedded homogeneity of beams in the nonlinear domain.
\newblock {\em International Journal of Solids and Structures}, 58:201--206,
  2015.

\bibitem{ericksen}
J~L Ericksen.
\newblock Uniformity in shells.
\newblock {\em Archive for Rational Mechanics and Analysis}, 77:73--84, 1970.

\bibitem{greennaghdi}
A~E Green and P~M Naghdi.
\newblock Non-isothermal theory of rods, plates, and shells.
\newblock {\em International Journal of Solids and Structures}, 6:209--244,
  1970.

\bibitem{koiter}
W~T Koiter.
\newblock On the nonlinear theory of thin elastic shells.
\newblock {\em Proc Knonklijke Nederlandse Akademie van Wetenschappen},
  B69:1--54, 1966.

\bibitem{kondo2}
K~Kondo.
\newblock Non-{R}iemannian geometry of imperfect crystals from a macroscopic
  viewpoint.
\newblock {\em RAAG Memoirs}, 1:458--469, 1955.

\bibitem{kroner81}
E~Kr{\"o}ner.
\newblock Continuum theory of defects.
\newblock In R~Balian et~al., editor, {\em Les Houches, Session XXXV, 1980 --
  Physique des d{\'e}fauts}, pages 215--315. North-Holland, New York, 1981.

\bibitem{lee}
J~M Lee.
\newblock {\em Manifolds and differential geometry}.
\newblock American Mathematical Society, Providence, Rhode Island, 2012.

\bibitem{malcolm}
D~J Malcolm and P~G Glockner.
\newblock Nonlinear sandwich shell and {C}osserat surface theory.
\newblock {\em Journal of the Engineering Mechanics Division}, 98:1183--1203,
  1972.

\bibitem{noll}
W~Noll.
\newblock Materially uniform bodies with inhomogeneities.
\newblock {\em Archive for Rational Mechanics and Analysis}, 27:1--32, 1967.

\bibitem{reissner}
E~Reissner.
\newblock Linear and nonlinear theory of shells.
\newblock In Y~C Fung and E~E Sechler, editors, {\em Thin-shell structures:
  Theory, experiment, and design}, pages 29--44. Prentice-Hall, Inc. Englewood
  Cliffs, New Jersey, 1974.

\bibitem{steigmann14}
D~J Steigmann.
\newblock Mechanics of materially uniform thin films.
\newblock {\em Mathematics and Mechanics of Solids}, 20:309--326, 2015.

\bibitem{steigmann99}
D~J Steigmann and R~W Ogden.
\newblock Elastic surface-substrate interactions.
\newblock {\em Proceedings of the Royal Society at London A}, 455:437--474,
  1999.

\bibitem{wangshell}
C-C Wang.
\newblock Material uniformity and homogeneity in shells.
\newblock {\em Archive for Rational Mechanics and Analysis}, 47:343--368, 1972.

\bibitem{arash}
A~Yavari and A~Goriely.
\newblock {R}iemann-{C}artan geometry of nonlinear dislocation mechanics.
\newblock {\em Archive for Rational Mechanics and Analysis}, 205:59--118, 2012.

\bibitem{zubov1}
L~M Zubov.
\newblock Von {K}\'arm\'an equations for an elastic plate with dislocations and
  disclinations.
\newblock {\em Doklady Physics}, 52:67--70, 2007.

\bibitem{zubov2}
L~M Zubov.
\newblock The linear theory of dislocations and disclinations in elastic
  shells.
\newblock {\em Journal of Applied Mathematics and Mechanics}, 74:63--72, 2010.

\end{thebibliography}

\end{document}